\documentclass[aps,prb,preprint,superscriptaddress,longbibliography]{revtex4-2}

\usepackage[pdftex]{graphicx}
\usepackage{subfigure}
\usepackage{hyperref}
\usepackage{units}
\usepackage{amsmath}
\usepackage{color}
\usepackage{amsfonts}

\begin{document}

\title{Coherent heavy charge carriers in an organic conductor near the bandwidth-controlled Mott transition}

\author{S. Oberbauer}
\altaffiliation[Present address: ]{attocube systems AG, 85540 Haar, Germany}
\affiliation{Walther-Mei{\ss}ner-Institut, Bayerische Akademie der Wissenschaften, Walther-Mei{\ss}ner-Strasse 8, D-85748 Garching, Germany}
\affiliation{Physik-Department, Technische Universit{\"a}t M{\"u}nchen, D-$85748$ Garching, Germany}

\author{S. Erkenov}
\affiliation{Walther-Mei{\ss}ner-Institut, Bayerische Akademie der Wissenschaften, Walther-Mei{\ss}ner-Strasse 8, D-85748 Garching, Germany}
\affiliation{Physik-Department, Technische Universit{\"a}t M{\"u}nchen, D-$85748$ Garching, Germany}

\author{W. Biberacher}
\affiliation{Walther-Mei{\ss}ner-Institut, Bayerische Akademie der Wissenschaften, Walther-Mei{\ss}ner-Strasse 8, D-85748 Garching, Germany}

\author{N. D. Kushch}
\affiliation{Walther-Mei{\ss}ner-Institut, Bayerische Akademie der Wissenschaften, Walther-Mei{\ss}ner-Strasse 8, D-85748 Garching, Germany}
\affiliation{Institute of Problems of Chemical Physics, Russian Academy of Sciences,  Ac. Semenov avenue 1, Chernogolovka, 142432 Russian Federation}

\author{R. Gross}
\affiliation{Walther-Mei{\ss}ner-Institut, Bayerische Akademie der Wissenschaften, Walther-Mei{\ss}ner-Strasse 8, D-85748 Garching, Germany}
\affiliation{Physik-Department, Technische Universit{\"a}t M{\"u}nchen, D-$85748$ Garching, Germany}
\affiliation{Munich Center for Quantum Science and Technology (MCQST), D-80799 Munich, Germany}

\author{M. V. Kartsovnik}
\email{mark.kartsovnik@wmi.badw.de}
\affiliation{Walther-Mei{\ss}ner-Institut, Bayerische Akademie der Wissenschaften, Walther-Mei{\ss}ner-Strasse 8, D-85748 Garching, Germany}

\date{\today}

\begin{abstract}
The physics of the Mott metal-insulator transition (MIT) has attracted huge interest in the last decades. However, despite broad efforts, some key theoretical predictions are still lacking experimental confirmation. In particular, it is not clear whether the large coherent Fermi surface survives in immediate proximity to the bandwidth-controlled first-order MIT. A quantitative experimental verification of the predicted behavior of the quasiparticle effective mass, renormalized by many-body interactions, is also missing. Here we address these issues by employing organic $\kappa$-type salts as exemplary quasi-two-dimensional bandwidth-controlled Mott insulators and gaining direct access to their charge carrier properties via magnetic quantum oscillations. We trace the evolution of the effective cyclotron mass as the conduction bandwidth is tuned very close to the MIT by means of precisely controlled external pressure. We find that the sensitivity of the mass renormalization to tiny changes of the bandwidth is significantly stronger than theoretically predicted and is even further enhanced upon entering the transition region where the metallic and insulating phases coexist. On the other hand, even on the very edge of its existence, the metallic ground state preserves a large coherent Fermi surface with no significant enhancement of scattering.
\end{abstract}

\maketitle

\section{Introduction}
Most of the popular ``bad metals'' exhibiting a Mott-insulating ground state are rather complex materials with more than one conduction band and various competing electronic and structural instabilities of the normal state \cite{imad98,hans13,lee06}. This renders an explicit quantitative comparison between experimental data and theory extremely difficult if not impossible. Moreover, in most materials it is very difficult to efficiently and controllably tune the electronic correlations within one and the same sample.
In this context layered organic conductors \cite{ishi98,toyo07} have several decisive advantages.
Firstly, they have simple quasi-two-dimensional (quasi-2D) electronic band structures.
Secondly, the invlolved relatively low energy scales and high compressibility allow us to access different regions of the phase diagram with a single sample by applying moderate
pressure in the 1\,GPa range, sometimes even below 100\,MPa.
Thirdly, the single crystals of organic conductors are usually intrinsically clean and homogeneous, which is particularly important for studies on the border of phase stability.

\begin{figure}[tb]
\center
\includegraphics[width = 0.6 \columnwidth]{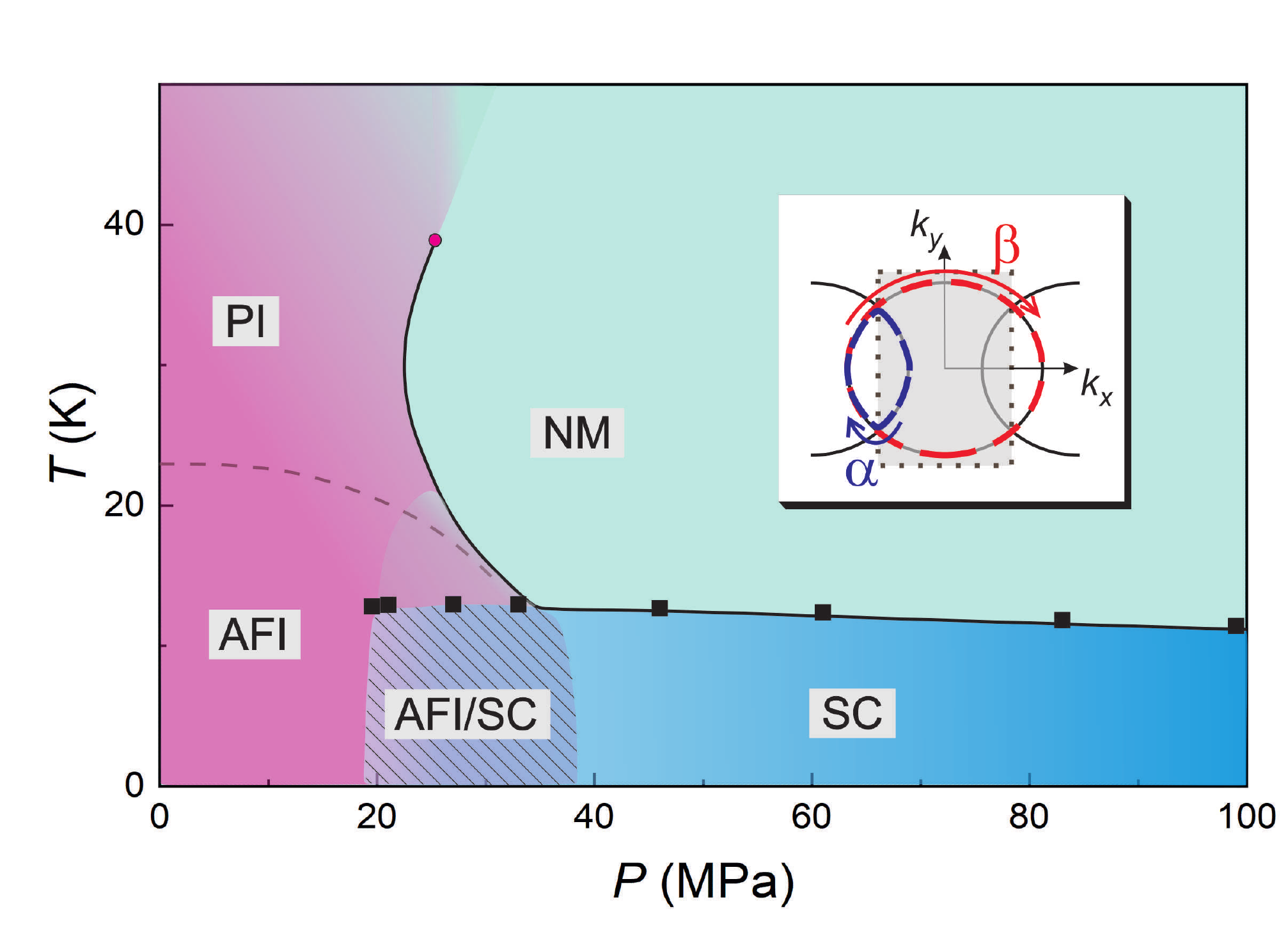}
\caption{Pressure--temperature phase diagram of $\kappa$-Cl based on the data from Refs. \citenum{lefe00,kaga04}). Black squares are the SC transition temperatures obtained from our zero-field resistive measurements, see inset in Fig.\,\ref{R-B}(b). Inset: 2D Fermi surface (thick solid lines) of $\kappa$-Cl in the NM state; the gray rectangle is the first Brillouin zone. Dashed lines indicate the classical cyclotron orbit $\alpha$ (blue) and the magnetic-breakdown orbit $\beta$ (red) in a strong magnetic field \cite{kart95c,yama96}, see text.}
\label{PhaseDia}
\end{figure}

The salts $\kappa$-(BEDT-TTF)$_2$X [where BEDT-TTF is the organic donor molecule bis\-(ethylene\-di\-thio)\-tetra\-thia\-fulvalene and X$^-$ a monovalent anion]
are materials with an effectively half-filled conduction band and
anisotropic triangular lattices. They
have long been known as model systems for studying the bandwidth-controlled quasi-2D Mott instability and closely related other fascinating phenomena such as unconventional superconductivity, quantum spin-liquid or valence-bond-solid states \cite{lefe00,kaga04,kano04,powe06,arda12,wosn19,powe11,kano11,ried19,pust22,miks21}.
The salt with X = Cu[N(CN)$_2$]Cl (hereafter $\kappa$-Cl) exhibits an antiferromagnetic Mott-insulating (AFI) ground state.
The material is very sensitive to pressure: already at pressures between 20 and 40\, MPa it undergoes a hysteretic first-order transition from the AFI (or, at $T > 20$ K, from the paramagnetic insulating, PI) state to the normal metallic (NM) and superconducting (SC) states \cite{lefe00,kaga04}, see Fig.\,\ref{PhaseDia}.
A similar result can be achieved by minor chemical modifications, which are generally supposed to affect the conduction bandwidth and therefore are often regarded as ``chemical pressure'' \cite{mori99a,kano04,powe06,arda12,meri08,pust18,sait21}.
As opposed to $\kappa$-Cl, the salts with X\,= Cu(NCS)$_2$ and Cu[N(CN)$_2$]Br (abbreviated as $\kappa$-NCS and $\kappa$-Br, respectively) are metallic and superconducting already at ambient pressure. The metallic state is characterized by the 2D Fermi surface shown in the inset in  Fig.\,\ref{PhaseDia}  (thick black lines) along with the first Brillouin zone (gray rectangle). The colored dashed lines and arrows in the inset indicate cyclotron orbits in a strong magnetic field \cite{kart95c,yama96}: a classical orbit $\alpha$ (blue) on the pocket centered on the border of the Brillouin zone and a magnetic-breakdown orbit $\beta$ (red), which encompasses the entire Fermi surface, with the area equal to the Brillouin zone area.

Presently, we possess ample information on the phase diagram as well as on the properties of the individual phases of the $\kappa$ salts \cite{ishi98,toyo07,kano04,powe06,arda12,wosn19,powe11,kano11,lefe00,kaga04,ried19,pust22,miks21,mori99a,meri08,pust18,sait21,kart95c,yama96,hart14,souz07,lunk12,lunk15,gati16,muel17,dres18}. What is, however, missing is the understanding of how the metallic ground state is changing in the very vicinity of the Mott metal-insulator transition (MIT).
For example, some theoretical works proposed a pseudogap formation in the $\kappa$ salts \cite{parc04,kang11,meri14,tana19}, whereas others argued against the pseudogap formation in exactly half-filled band systems \cite{hebe15,brag18}
such as our materials.
Moreover, a dramatic increase of the scattering rate to values comparable to the nearest-neighbor hopping rate was predicted for the transitional region of the phase diagram where the metallic phase coexists with the insulating one \cite{park08}. In this situation, the very existence of coherent Landau quasiparticles with a well-defined Fermi surface comes into question.
From the experimental side, some indirect hints toward a pseudogap formation
were found in the NMR and Nernst-effect studies of the $\kappa$-Br salt located on the metallic side near the MIT \cite{miya02,nam07}. However, a decisive test probing the evolution of the Fermi surface at a controlled variation of the correlation strength has been lacking.

Another important unsolved issue concerns the behavior of the effective mass $m$ of the charge carriers.
The specific heat measurements on the partially deuterated  $\kappa$-Br samples \cite{naka00} suggested a significant decrease of the effective mass at approaching the MIT.
This result apparently conflicts with theoretical predictions about the interaction-induced enhancement of the mass \cite{brin70,geor96}.
On the other hand, in qualitative agreement with the theory, infrared experiments on mixed $\kappa$-Cl/Br crystals \cite{meri08} revealed a higher mass for the
crystal with the higher Cl content, hence more close to the insulating state. However, a quantitative comparison cannot be done due to the restricted amount of data
(only two compositions have been analyzed) and uncertainty in the location on the phase diagram.

Here we address the above issues by studying the Shubnikov-de Haas (SdH) oscillations in the pressurized $\kappa$-Cl salt. Magnetic quantum oscillations have proved extremely useful for characterizing the conduction system of layered organic metals \cite{wosn96,kart04} as well as in other correlated-electron materials of topical interest such as cuprate \cite{seba15,helm15,chan16} and iron-based \cite{carr11,cold18,tera18} superconductors, topological conductors \cite{moll16,pezz18,wang21}, and heavy-fermion compounds \cite{jiao15,xian18,bast19}. In contrast to most thermodynamic methods collecting an integrated information from the whole bulk of the sample, SdH oscillations are a selective probe of the metallic state. This is particularly important for exploring the inhomogeneous region of the first-order MIT, where the metallic phase is intertwined with the insulating one.

\section{Methods}
{\it Experimental --} The samples studied in this work are single crystals of $\kappa$-Cl and $\kappa$-NCS, grown electrochemically according to literature
\cite{ishi98,will90,uray88},
with a lateral size of about $0.5 \times 0.5 \mathrm{\,mm}^2$ and thickness (along the least conducting direction) of $0.05-0.3 \mathrm{\,mm}$. Several batches were screened to select high-quality samples, by measuring $T$- and $B$-dependent resistance, see Supplemental Material (SM) below for details.

\begin{figure}[tb]
\center
\includegraphics[width = 0.5 \columnwidth]{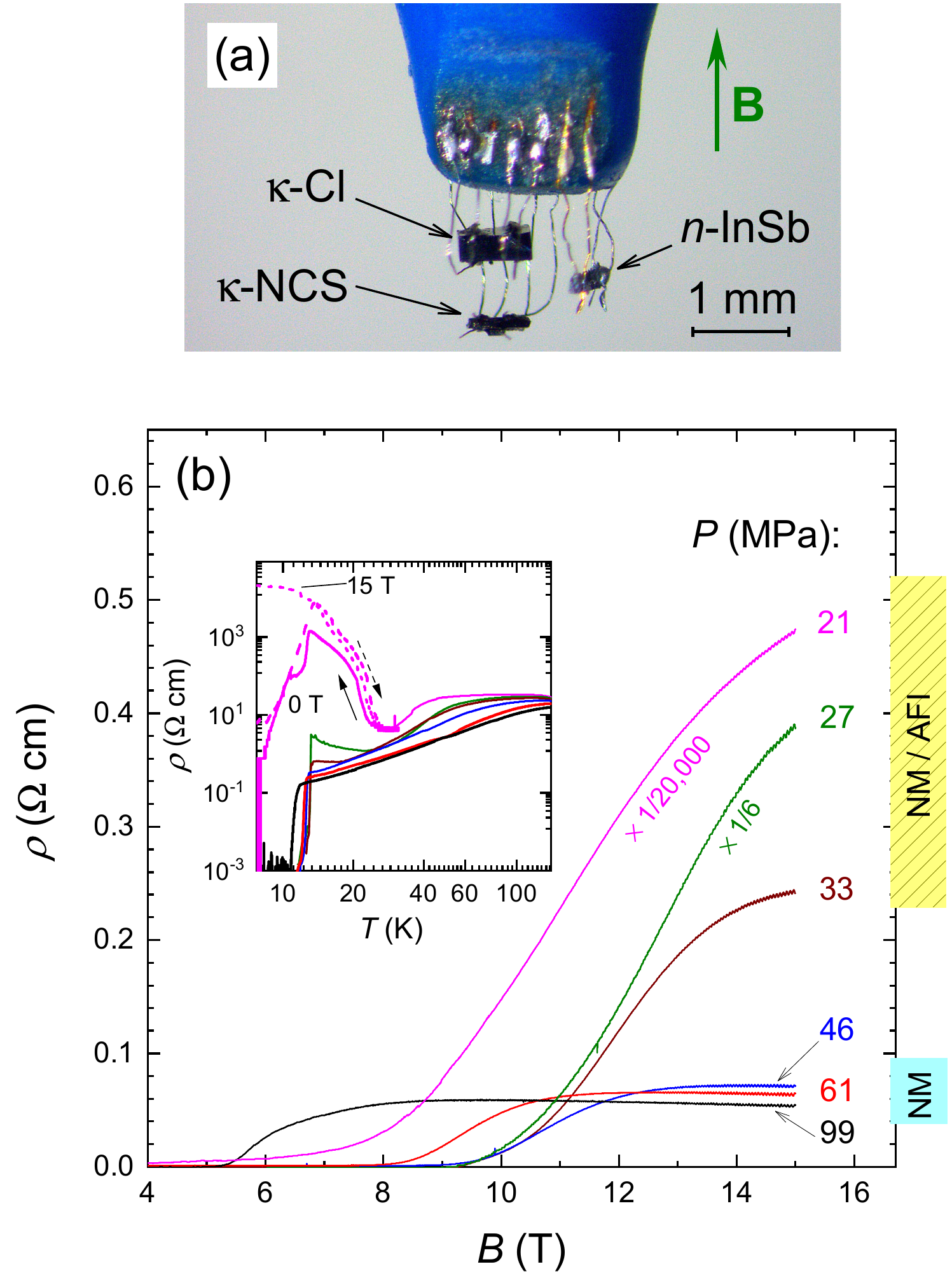}
\caption{(a) Crystals of $\kappa$-Cl (sample \#1) and $\kappa$-NCS crystals along with the InSb pressure sensor mounted for resistive measurements in the pressure cell,  before loading into the cell. The samples are aligned for measurements in a magnetic field $\mathbf{B}$ perpendicular to the layers.
(b) Examples of the field-dependent resistivity of $\kappa$-Cl sample \#1 at different pressures, at $T=100$\,mK. Pressures $P\leq 33$\,MPa correspond to the NM/AFI coexistence region of the phase diagram. The curves for $P = 27$ and $21$\,MPa are scaled down by a factor of 6 and 20,000, respectively. The inset shows the temperature dependence of the zero-field resistivity at the same pressures.
For $P=21$\,MPa, the solid and dashed lines correspond to the down and up $T$-sweeps, respectively, revealing a strong hysteresis in the phase-coexistence regime. In addition, a $T$-sweep up under a magnetic field of 15\,T, suppressing the SC transition,  is shown for this pressure (dotted line).}
\label{R-B}
\end{figure}

The interlayer resistance of the samples was measured using  the standard 4-probe low-frequency ac technique, in magnetic fields of up to 15\,T applied perpendicular to the layers. The samples were attached to 20\,$\mu$m-thick annealed Pt wires serving as electrical leads [see Fig.\,\ref{R-B}(a)] in a small piston-cylinder cell made of pure BeCu. The silicone oil GKZh \cite{bura08} was used as a pressure medium. Pressure was applied at room temperature and monitored by measuring the resistance of a calibrated pressure sensor placed side by side with the samples, see Fig.\,\ref{R-B}(a) and SM. For precise determination of pressures below 100\,MPa, an $n$-doped InSb sensor \cite{konc78} was adapted with the sensitivity  $\frac{1}{R}\frac{dR}{dP} \approx 0.35$\,GPa$^{-1}$. With this,
we were able to measure pressure with an accuracy of $\pm 2$\,MPa. The $P$ values given in the text are those measured at $T = 15$\,K.

{\it Evaluation of $m_c$ and $T_{\mathrm{D}}$ --} The cyclotron mass $m_c$ was evaluated from the temperature dependence of the SdH amplitude by fitting the latter with the standard Lifshitz-Kosevich
temperature damping factor \cite{shoe84} $R_T = \frac{Km_cT/B}{\sinh(Km_cT/B)}$, where
$K=2\pi^2k_{\mathrm{B}}/\hbar e$ with $k_{\mathrm{B}}$
being the Boltzmann constant and $e$ the elementary charge. The mass was determined for $\kappa$-Cl samples \#1 and\#2 studied in the $T$ intervals $0.1 - 0.8$\,K and $0.45-1.0$\,K, respectively, and for the $\kappa$-NCS crystal studied simultaneously with $\kappa$-Cl sample \#1.
Details of evaluations are given in Secs.\,S-III and S-IV of SM.

The Dingle temperature is commonly extracted from the field dependence of the SdH amplitude described by the Dingle factor, $R_{\mathrm{D}} = \exp(-Km_cT_{\mathrm{D}}/B)$. In our case, a potential complication might arise due the magnetic-breakdown origin of the oscillations analyzed in this work, see Sec.\,\ref{MR} for the description of the oscillations.
However, our estimation of the breakdown field yields a very low value, $B_0\simeq 1$\,T, see Sec.\,S-II of SM . Therefore, for our field range of interest, $12 \leq B \leq 15$\,T, the corresponding magnetic-breakdown factor in the oscillation amplitude is $R_{\mathrm{MB}} = \exp(-2B_0/B) \simeq 1$ and its variation with field ($\leq 2.5\%$) can be neglected in comparison to that of $R_{\mathrm{D}}$.
Another complication in the field dependence comes from a low-frequency beating of the oscillations. Beats of quantum oscillations are often observed in the layered organics and associated with a weak warping of the Fermi surface cylinder \cite{wosn96,kart04}. Our analysis of the field dependence taking into account the beats and yielding the Dingle temperature is presented in detail in Sec.\,S-V of SM.

\section{Results and Discussion}\label{Results}
\subsection{Resistive behavior and Fermi surface near and at the MIT}\label{MR}

The overall behavior of the interlayer resistivity of pressurized $\kappa$-Cl is illustrated in Fig.\,\ref{R-B}. The displayed pressure range, 20 to 100\,MPa, includes both the purely metallic region of the phase diagram and the transitional region, where the metallic and insulating phases coexist.
The phase-coexistence regime is readily detected due to a strong enhancement of the resistivity at temperatures between 25\,K and the SC transition temperature $T_c \simeq 12.5$\,K (inset in Fig.\,\ref{R-B}). At lower temperatures this enhancement is hidden by the SC transition but becomes obvious when superconductivity is suppressed by magnetic field.

\begin{figure}[tb]
\center
\includegraphics[width = 0.6 \columnwidth]{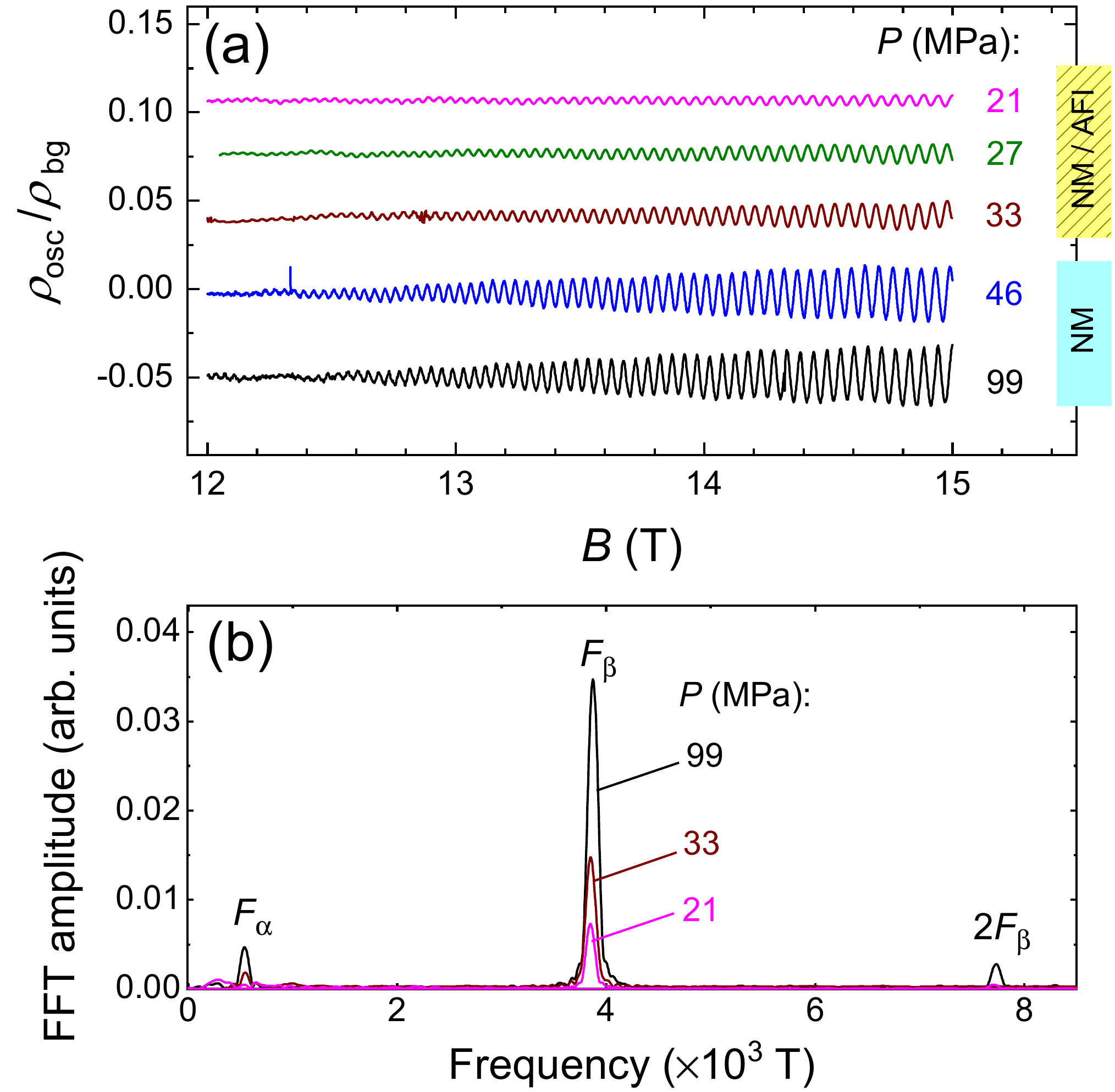}
\caption{(a) Oscillatory component of resistivity $\rho_{\text{osc}}(B) \equiv \rho(B)-\rho_{\text{bg}}(B)$ in $\kappa$-Cl, normalized to the nonoscillating resistivity background $\rho_{\text{bg}}(B)$; data obtained at different pressures, at $T = 100$\,mK. The curves are vertically shifted for clarity. (b) FFT spectra for three curves from panel (a). The dominant peak originates from the magnetic-breakdown orbit $\beta$ encircling the entire Fermi surface, see inset in Fig.\,\ref{PhaseDia}. Its persistence at the lowest pressure indicates that the large Fermi surface survives, without a significant change, inside the coexistence region, even very close to the purely insulating region of the phase diagram.
}
\label{Oscis}
\end{figure}

The SdH oscillations are found at all pressures, at $B > 10$\,T and $T\leq 1$\,K. Examples of the oscillatory component of the resistivity
are shown in Fig.\,\ref{Oscis}(a).

Note that sizeable oscillations are observed even at $P$ very close to $P_{c1}\simeq 20$\,MPa, the lower border of the coexistence region. This is surprising: at this pressure the metallic phase occupies only a tiny fraction of the sample. In fact, this fraction is
far below the standard percolation limit, as inferred from the NMR measurements \cite{lefe00} and reflected in a dramatic, five orders of magnitude, increase of the measured normal-state resistivity, see Fig.\,\ref{R-B}.
Keeping in mind that SdH oscillations are a fingerprint of a well-defined Fermi surface, our observation provides firm evidence of a narrow continuous path for coherent charge transport even at the very edge of the existence of the metallic phase.

Figure \ref{Oscis}(b) shows examples of the fast Fourier transform (FFT) of the oscillatory signal.
The dominant peak at frequency $F_{\beta} \approx 3850$\,T is associated with the large magnetic-breakdown orbit $\beta$ encompassing the entire 2D Fermi surface and having an area equal to that of the first Brillouin zone
(see inset in Fig.\,\ref{PhaseDia}). Importantly, this peak, found earlier at high pressures in the purely NM state \cite{kart95c,yama96}, persists without a notable shift in the coexistence region. This means that the Fermi surface of the metallic phase remains largely unchanged upon entering the metastable phase-coexistence regime.

In addition to the main frequency $F_{\beta}$ and its weak second harmonic, a low SdH frequency $F_{\alpha} \approx 540$\,T, originating from the classical orbit $\alpha$ \cite{kart95c,yama96} on the closed Fermi pocket centered at the Brillouin zone boundary (blue loop in the inset to Fig.\,\ref{PhaseDia}), is detected in Fig.\,\ref{Oscis}.
The dominant contribution of the magnetic-breakdown oscillations $\beta$
implies that there is only a small gap between the open and closed portions of the Fermi surface, $\Delta_{\text{MB}} \simeq 1$\,meV.
This is consistent with the theoretical estimation \cite{wint17}. In the following we focus on these $\beta$ oscillations, which probe the properties of the entire Fermi surface.

\subsection{Effective cyclotron mass}\label{mc}
\begin{figure}[tb]
\center
\includegraphics[width = 0.6 \columnwidth]{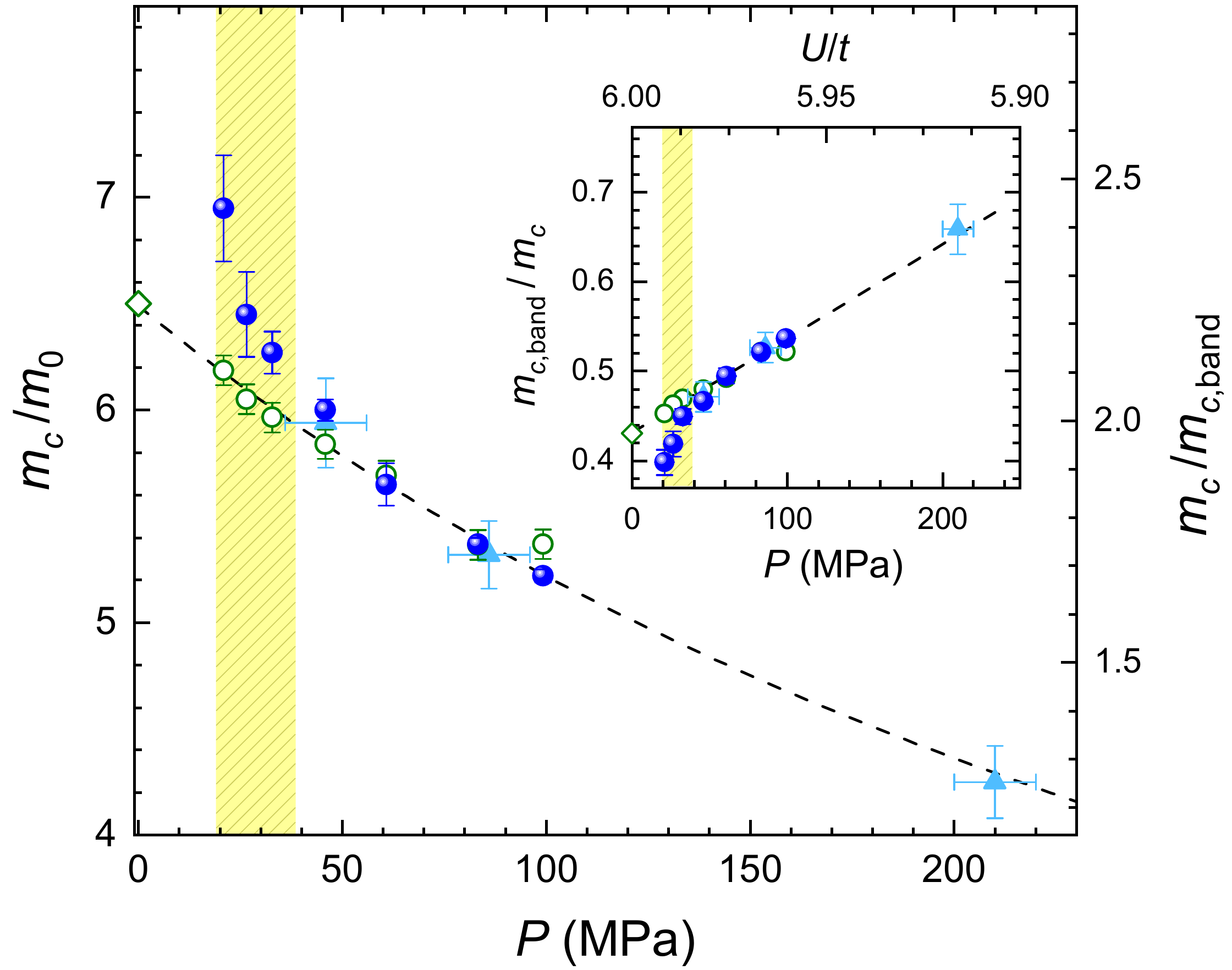}
\caption{Pressure-dependent effective cyclotron mass of $\kappa$-Cl samples \#1 (solid circles) and \#2 (triangles) along with the mass obtained for the $\kappa$-NCS crystal (open circles) measured in parallel with $\kappa$-Cl sample \#1. Note that $\kappa$-NCS remains metallic down to ambient pressure; the corresponding mass value at $P=0$ (diamond) is taken from Ref. \citenum{caul94}. The left scale is given in units of the free electron mass $m_0$, the right scale in units of the band cyclotron mass $m_{c,\text{band}} = 2.8m_0$ \cite{xu95,meri00a}. Both materials exhibit the same inverse-linear pressure dependence in the homogeneous NM state; the dashed line is the fit to Eq.\,(\ref{mass}). The hatched rectangle shows the phase coexistence region for the $\kappa$-Cl salt. Inset: inverse renormalization factor $m_{c,\text{band}}/m_c$ demonstrating the $P$-linear dependence in the purely metallic state and the deviation of $\kappa$-Cl from this behavior in the coexistence region. The top-axis scale shows the $U/t$ ratio estimated based on the band structure calculations \cite{kand09}, see text.
}
\label{m-P}
\end{figure}

Further insight into the conduction system is gained from the Lifshitz-Kosevich analysis of the SdH amplitude, which yields the effective cyclotron mass $m_c$ of the charge carriers from the temperature damping factor $R_T$ \cite{shoe84}.
The ratio of the $m_c$ to the band cyclotron mass $m_{c,\text{band}}$, obtained from one-electron band structure calculations, is determined by many-body interactions. In the vicinity of the MIT, it provides a quantitative measure of the electronic interaction strength \cite{brin70,geor96}:
the inverse renormalization factor, $m_{c,\text{band}}/m_{c}$, is usually equated to the quasiparticle residue $Z$, a key parameter of the Fermi-liquid theory.
In Fig.\,\ref{m-P} we plot our results on the pressure-dependent cyclotron mass;
solid symbols represent two $\kappa$-Cl samples (see SM for details of the data analysis).

In order to better elucidate the role of the proximity to the MIT, we confront the $\kappa$-Cl salt  with the sister compound $\kappa$-NCS.
To this end, we have measured SdH oscillations on a $\kappa$-NCS crystal simultaneously with
the $\kappa$-Cl sample $\#1$, under the same conditions.
Both materials have very similar quasi-2D electronic band structures and Fermi surfaces.
In particular, the sizes of the $\beta$ orbit differ by less than $1\%$, see Sec S-IV of SM.
The essential difference, however, is that $\kappa$-NCS is metallic and superconducting already at ambient pressure.
Although the amplitude of the $\beta$ oscillations in $\kappa$-NCS is relatively weak due to a larger, than in $\kappa$-Cl, magnetic-breakdown gap, we succeeded in evaluating the relevant effective mass at all pressures applied (see SM for details).
The results are presented in Fig.\,\ref{m-P} by open circles.

Two important observations can be drawn immediately from a glance at the data in Fig.\,\ref{m-P}. Firstly, both materials show a significant increase of the cyclotron mass at decreasing $P$ -- a clear manifestation of the growing electronic correlations in the vicinity of the MIT. Secondly, the masses of the two salts are practically indistinguishable at $P>40$\,MPa, but start to diverge from each other at lower pressures, where $\kappa$-Cl enters the NM/AFI coexistence region (shaded area in Fig.\,\ref{m-P}).

To further elaborate the first observation, we note that within the homogeneous NM part of the phase diagram, the experimental pressure dependence of the mass is well described by the simple expression:
\begin{equation}\label{mass}
 m_c \propto \left( P - P_0 \right)^{-1}\,
\end{equation}
with $P_0 = (-410 \pm 20)$\,MPa for both salts (dashed lines in Fig.\,\ref{m-P}).
This result looks fully consistent with theoretical predictions.
Indeed, both the Brinkman-Rice theory \cite{brin70} and the (single-site) dynamical mean-field theory (DMFT) \cite{geor96} predict a variation of the renormalized effective mass in the form:
\begin{equation}\label{BR}
\frac{m_{c,\text{band}}}{m_c} = Z \approx C_Z\left[1-\left(\frac{U/t}{(U/t)_0} \right)  \right]
\end{equation}
in close proximity to the MIT\footnote{Strictly speaking, the theory \cite{brin70,geor96} usually considers the conventional quasiparticle effective mass $m = \hbar^2\left( \frac{\partial^2 E}{\partial k^2} \right)^{-1}$. However, the very same many-body renormalization factor is also applied to the cyclotron mass $m_c$ entering the expressions for the magnetic quantum oscillations \cite{shoe84}.}.
Here the correlation strength is quantified by the ratio between the effective on-site Coulomb repulsion $U$ and the nearest-neighbor transfer integral $t$
and the prefactor $C_Z = 2$ and $\approx 0.9$ in the Brinkman-Rice theory \cite{brin70}  and DMFT \cite{geor96}, respectively.
The formal divergence of the mass at pressure $P_0$, determined by the critical correlation strength ratio $(U/t)_0 \simeq 12$,
is cut off at the first-order MIT \cite{geor96}.
More advanced theories \cite{park08,balz09,scha15,loon18,rohr18,wyso17} show that magnetic interactions in an anisotropic triangular lattice may further shift the transition to considerably lower $U/t$, that is, to higher pressures. However, they should not dramatically affect the shape of the dependence $Z(U/t) \propto 1/m_c(U/t)$. It is easy to see that the expressions (\ref{mass}) and (\ref{BR}) are equivalent once we assume that small pressure-induced changes of the correlation strength are linear in $P$:
\begin{equation}
1 - \frac{U/t}{(U/t)_0} = \alpha \left(  P - P_0 \right) \ll 1\,.
\label{U-P}
\end{equation}

The importance of magnetic interactions is clearly manifest in the fact that the AFI ground state of $\kappa$-Cl sets in already at positive pressures of about 40\,MPa, despite the negative $P_0$ in Eq.\,(\ref{mass}).
By contrast, the sister compound $\kappa$-NCS remains metallic down to ambient pressure, even though it shows
practically identical $m_c(P)$ values,
hence, the same correlation strength $U/t$ at the same pressures!
This is in fact a prominent demonstration of the geometrical spin frustration in a triangular lattice.
Indeed, first-principles band structure calculations \cite{kand09,kore14} show that the
ratio between the next-nearest- and nearest-neighbor transfer integrals, $t'/t$, characterizing the frustration ($t'/t = 1$ in the fully frustrated lattice),
is significantly higher in $\kappa$-NCS than in $\kappa$-Cl. A stronger frustration is expected to weaken antiferromagnetic correlations, thereby suppressing the insulating instability.
It is the frustration ratio $t'/t$ rather than the correlation strength ratio $U/t$, which is proposed to be responsible for the difference in the ground states of the $\kappa$-NCS and $\kappa$-Cl salts
 \cite{kore14}.
Our experiment, revealing the close similarity of the $U/t$ values in the two salts with different zero-pressure ground states, provides a firm support for this theoretical prediction.

As a next step, we aim at making a more quantitative comparison with theory by estimating the factor $C_Z$, which  characterizes the steepness of the $m_c(U/t)$ dependence  in Eq.\,(\ref{BR}).
To this end, we use the results of ab-initio band structure calculations performed for $\kappa$-NCS at two pressures yielding $U/t= 6.0$ and $5.7$ for $P = 0$ and $0.75$\,GPa, respectively \cite{kand09}. We further
assume that the linear relationship between $U/t$ and $P$, given by Eq.\,(\ref{U-P}), holds throughout this pressure range.
The top axis in the inset in Fig.\,\ref{m-P} presents the resulting $U/t$ scale.
The linear fit to the dependence $m_{c,\text{band}}/m_c(U/t)$ yields $C_Z = 16.0 \pm 0.4$. This is an order of magnitude larger than the theoretical values.

Of course, our estimation, based on a two-point linear interpolation of the $U/t(P)$ dependence in a rather large pressure range, is not very precise. An additional uncertainty stems from the band structure calculations whose results are quite sensitive to the model and calculation method used, cf. Refs.\,\citenum{kand09,kore14,naka09,naka12}. A detailed targeted calculation for our salts in the pressure range below 100\,MPa, especially taking into account the many-body effects, would be very helpful for reducing the error bar.
Further on,
magnetic interactions, a variation of the frustration ratio $t'/t(P)$, and electron-phonon coupling may contribute to the variation of the mass renormalization.
For example, one may consider scattering on spin fluctuations as a source of additional mass renormalization. Such an effect is observed near antiferromagnetic quantum phase transitions in some heavy-fermion materials \cite{gege08,mats11,sett07} as well as in pnictide \cite{walm13} and possibly cuprate \cite{seba10a,helm15} superconductors. A critical increase of scattering suppresses the quasiparticle spectral weight $Z$ around the ``hot spots'' of the Fermi surface connected by the antiferromagnetic wave vector, thereby inducing a pseudogap \cite{gege08,kang11,meri14} and leading to a divergence of the effective mass \cite{rama01}. This happens, however, only in narrow areas around the hot spots while the rest of the Fermi surface is unaffected. Therefore, the cyclotron mass, an integral characteristic of the entire cyclotron orbit on the Fermi surface, is only moderately increased. By contrast to the abovementioned materials exhibiting an antiferromagnetic quantum criticality, our compounds undergo a first order transition driven by electronic correlations. Magnetic interactions are a secondary effect. They shift the transition to lower $U/t$, but are not expected to change the behavior of $Z(U/t)$ in the metallic phase \cite{park08,wata06,powe05}. As shown in Sec.\,\ref{TD}, the pseudogap associated with scattering on spin fluctuations, if exists, does not exceed few millielectronvolt. Therefore, a significant contribution of magnetic interactions to the mass enhancement is unlikely. In general, any significant contribution, besides the electron-electron correlations, to the mass renormalization would lead to a violation of the simple relationship in Eq.\,(\ref{BR}), which is obviously not the case. Thus, it is highly unlikely that the mentioned factors might account for the drastic, order-of-magnitude enhancement of $C_Z$.

Due to the lack of other systematic experimental data, it is difficult to judge whether the observed dramatic quantitative discrepancy is specific to our compounds or has a more universal and fundamental origin. The few relevant experiments we are aware of are very recent infrared studies of two organic charge-transfer salts $\beta'$-EtMe$_3$Sb[Pd(dmit)$_2$]$_2$ \cite{li19} and $\kappa$-[(BEDT-STF)$_x$(BEDT-TTF)$_{1-x}$]$_2$Cu$_2$(CN)$_3$ \cite{pust21} and the early heat-capacity experiment on one of the best-studied inorganic Mott compounds V$_2$O$_3$ \cite{cart93}. A closer look at this data (Sec. S-V of SM) seems to point towards a general character of the present result. However, more work is needed for a decisive conclusion.

As noted above, the $m_c(P)$ curves in Fig.\,\ref{m-P} diverge from each other at low $P$: for the fully metallic $\kappa$-NCS the data perfectly obeys Eq.\,(\ref{mass}) down to ambient pressure, whereas $\kappa$-Cl displays a much steeper increase of $m_c$
at $P < 40$\,MPa.
This threshold remarkably coincides with the nucleation of the insulating phase upon entering the transition region of the phase diagram.
It is therefore tempting to link the
steepening of $m_c(P)$
with the coexistence of the NM and AFI phases. A trivial mechanism, which might alter the $P$-dependent mass in an inhomogeneous phase-separated environment, is due to internal strains. However, the formation of a domain with the lower-pressure, hence, lower-density AFI phase upon approaching the MIT from the metallic side can only increase the effective internal pressure in the adjacent NM domain. This should lead to an apparent flattening of the $P$-dependence, contrary to the experimentally observed acceleration.
This acceleration resembles to some extent the sharp enhancement of the $T$-dependent effective mass in vanadium dioxide in the phase-coexistence region of the thermally-driven first-order MIT inferred from optical conductivity measurements \cite{qazi07}. In that compound it was associated with a prominent, $\approx 1000$\,cm$^{-1}$, pseudogap caused by fluctuating charge order. As argued above, it is highly unlikely that a fluctuating order leads to a noticeable contribution to the mass renormalization in our compounds.
An interesting scenario, which may be relevant to the observed behavior, is the formation of unusually thick domain walls at the first-order MIT, as was very recently proposed for a strongly frustrated half-filled Mott-Hubbard system \cite{suar21}. Such walls were argued to host poorly developed "resilient" quasiparticles with a spectral weight rapidly decreasing towards $T=0$ \cite{lee22}. The effective mass of these resilient quasiparticles is expected to diverge at the lower critical pressure (the border to the purely insulating state). While the theory \cite{suar21,lee22} has been done for an ideally frustrated lattice, one may expect that some traces of this effect persist in moderately frustrated systems like our materials. Still, there is a question how this should be reflected in the quantum oscillations as the latter would be then probing both the "good" quasiparticles in the purely metallic domains and the resilient quasiparticles in the domain walls. More work on materials with different degrees of frustration should be done for clarifying this problem.


\subsection{Scattering near the MIT: what can we learn from the SdH oscillations?}\label{TD}
Magnetic quantum oscillations, being critically sensitive to scattering, may be very helpful
for testing the quasiparticle coherence and the presence of a pseudogap near the MIT.
Qualitatively, the observation of the $\beta$ oscillations even deep inside the NM/AFI coexistence region appears to be by itself a signature of a large coherent Fermi surface.
On a quantitative level,
the effect of a finite quasiparticle lifetime $\tau$ on the oscillation amplitude is usually described in terms of the Dingle factor \cite{shoe84},
$R_{\text{D}} = \exp (-2\pi^2 k_{\text{B}}T_{\text{D}}/\hbar\omega_c)$,
where $k_{\text{B}}$ is the Boltzmann constant, $\omega_c = eB/m_c$ is the cyclotron frequency on the $\beta$ orbit, $e$ the elementary charge, and the Dingle temperature $T_{\text{D}} = \hbar/2\pi k_{\text{B}}\tau$.
A pseudogap, arising from the enhanced scattering of quasiparticles on short-range antiferromagnetic fluctuations, should effectively increase $T_{\text{D}}$.

\begin{figure}[tb]
\center
\includegraphics[width = 0.6 \columnwidth]{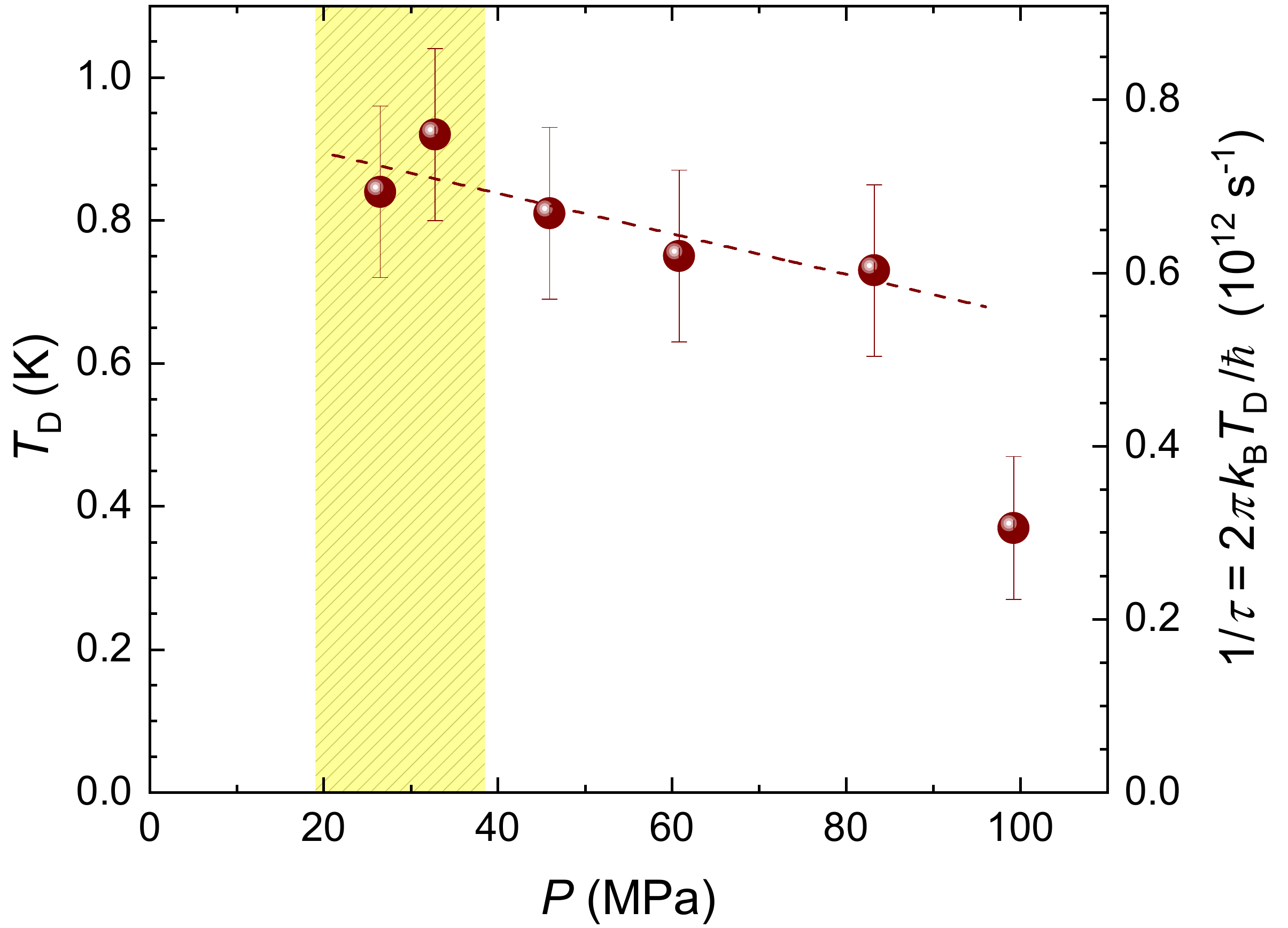}
\caption{Dingle temperature of $\kappa$-Cl versus pressure. No significant increase of $T_{\text{D}}$, hence no enhancement of scattering occurs upon entering the NM/AFI coexistence region (hatched zone). The dashed line is a linear fit to the data between 27 and 83\,MPa suggesting a weak gradual decrease of $T_{\text{D}}$ at increasing pressure. The right-hand axis shows re-scaling of the Dingle temperature to the scattering rate.}
\label{TD-P}
\end{figure}
In Fig.\,\ref{TD-P}, we plot the Dingle temperature for the $\kappa$-Cl sample $\#1$ evaluated from the $B$-dependent SdH amplitude at different pressures (see SM for details of the evaluation).
The obtained values, $T_{\text{D}} \lesssim 1$\,K (that is, $\tau \gtrsim 10^{-12}$\,s) are typical for clean crystals of organic metals \cite{wosn96,kart04}.
The sharp drop of $T_{\text{D}}$ near 100\,MPa was not reproduced in our additional test run, see SM, hence we disregard it as a spurious effect.
Taking the rest of the data in Fig.\,\ref{TD-P}, the Dingle temperature does not show a significant variation.
In particular, it is insensitive to the nucleation of the insulating phase below 40\,MPa. This excludes any  perceptible enhancement of scattering upon entering the coexistence regime, which one might expect based, e.g., on the cluster DMFT calculations \cite{park08}.

As a general trend, the data in Fig.\,\ref{TD-P}
seems to show slight enhancement of scattering at decreasing
pressure although the total change does not exceed the error bar of our evaluation.
An overall variation $\Delta T_{\text{D}} \approx 0.2$\,K within our pressure range
corresponds to the change in scattering rate $\delta(1/\tau) \approx 1.6\times 10^{11}$\,s$^{-1}$ or $0.1$\,meV in energy units. This is a  vanishingly small value as compared to the relevant energies near the MIT; in particular, it  is two orders of magnitude smaller than the anticipated pseudogap scale, $\delta_{\mathrm{PG}}\sim 20$\,meV \cite{meri14}.
It is important to note, however, that the standard Dingle model employed above
ignores the strong momentum dependence of scattering associated with the possible pseudogap formation.
As an opposite limiting case for the estimations, one may use an analogy with magnetic breakdown:
the electrons, travelling on the cyclotron orbit $\beta$ tunnel through two pairs of ``hot spots'' determined by the antiferromagnetic wave vector. The corresponding damping factor for the oscillation amplitude, $R_{\mathrm{MB}} = \exp(-2B_0/B)$, with a characteristic field $B_0 \simeq (m_c\delta_{\mathrm{PG}}/\hbar e)^2/F_{\beta}$,
has the same form of $B$-dependence as the Dingle factor. A straightforward rescaling of the $T_{\text{D}}$ variation in Fig.\,\ref{TD-P} yields a much larger upper estimate for the pseudogap, $\delta_{\mathrm{PG}} \lesssim 4$\,meV.
This value is still small but already closer to the theoretical predictions for the antiferromagnetic pseudogap.
However, one has to keep in mind that the magnetic breakdown scenario implies a full gap localized in a very narrow region of $\mathbf{k}$-space near the Fermi surface, whereas the pseudogap means a finite, though suppressed, quasiparticle density, spread over a relatively broad $\mathbf{k}$-space interval. Obviously, the true value of $\delta_{\mathrm{PG}}$ lies between the above two estimates.
For a more accurate evaluation we need an explicit inclusion of a pseudogap into the theoretical description of the oscillations. With such a theory at hand, the SdH oscillations will provide a powerful tool for accurate determination of the pseudogap.

Finally, it is worth noting that the data in Fig.\,\ref{TD-P} also sets a lower limit for the size of metallic domains in the coexistence region which obviously cannot be much smaller than the electron mean free path. The latter is evaluated as:
$\ell \sim \hbar k_{\text{F}}\tau/m_c \simeq \sqrt{2\hbar eF_{\beta}}\tau/m_c \approx 100$\,nm. This estimate is of course related to the plane of cyclotron orbits in magnetic field that is, the plane of conducting layers.
As to the interlayer direction, it was already noted in Sect.\,\ref{MR} that narrow coherent-transport channels persist throughout the sample
even at $P \simeq 20$\,MPa, at the very edge of the metallic state.
Thus, we conclude that the phase separation at the first-order MIT occurs on a macroscopic scale, exceeding the crystal lattice period by at least two orders of magnitude in all directions.

\section{Conclusions and outlook}
The high tunability of the electronic ground state and excellent crystal quality make the organic charge-transfer salts $\kappa$-(BEDT-TTF)$_2$X a perfect testbed for experimental exploration of the bandwidth-controlled Mott instability. These features were crucial for the success of our quantum-oscillation experiment on the pressurized $\kappa$-Cl and $\kappa$-NCS salts aimed at tracking the evolution of fundamental characteristics of the conducting system in the immediate proximity to the Mott-insulating state.

We have observed SdH oscillations throughout the entire pressure range studied, including the homogeneous metallic domain of the phase diagram close to the MIT as well as the transitional region, where the metallic and insulating phases coexist. At all pressures the dominant contribution to the oscillations comes from the magnetic-breakdown orbit $\beta$ with the area equal to that of the first Brillouin zone. This is a direct evidence that even at the very border of the purely insulating state, when just a minor fraction of the sample bulk is occupied by the metallic phase, the latter still preserves the large coherent Fermi surface, the same as far away from the MIT (cf. \cite{kart95c,yama96}).

The analysis of the field-dependent SdH amplitude shows that no notable enhancement of scattering occurs upon approaching the insulating phase. In particular, we find that the pseudogap associated with antiferromagnetic fluctuations, if it exists, does not exceed 4\,meV. A more definite and precise evaluation will be possible once a theoretical description of quantum oscillations in the presence of strong anisotropic scattering on magnetic fluctuations is available.

One of the main goals of this work was the precise determination of the pressure-dependent effective cyclotron mass near the MIT. Our data provides a firm experimental basis for an explicit quantitative test of theoretical predictions for the many-body renormalization effects in close proximity to the transition.
In the homogeneous metallic state the inverse effective mass is found to display a very simple $P$-linear behavior.
This simple functional dependence appears to be consistent with the expected renormalization effect caused by electron-electron interactions.
However---and this is perhaps the most intriguing result---the slope of this dependence turns out to be an order of magnitude steeper than expected and is even further accelerated upon entering the phase-coexistence region. The unexpectedly steep mass enhancement seems to be not just peculiar to the present two compounds. Verifying whether it indeed has a general character is a matter of further purposeful experiments on bandwidth-controlled Mott insulators with different strengths of magnetic interactions and different degrees of frustration, accompanied by rigorous band-structure calculations involving correlation effects. If this proves to be the case, it will demand a significant revision of our present understanding of the Mott transition physics.

Another interesting finding of this work is that the $\kappa$-Cl and $\kappa$-NCS salts exhibit equal mass-renormalization factors, hence, the same ($P$-dependent) $U/t$ values in the homogeneous NM state. This finding, along with the fact that the two compounds have different ground states at ambient pressure, provides a clear experimental evidence for the decisive role of geometrical spin frustration (ratio $t'/t$), which was predicted to be different in these salts \cite{kand09,kore14}. In this respect, the anion substitution, often referred to as ``chemical pressure'', acts differently from physical pressure.
It would be interesting to do similar SdH experiments on other $\kappa$ salts with different anions for elucidating the role of subtle chemical and structural modifications on electronic correlations near the MIT.

\acknowledgments
\noindent {\it Acknowledgments --} We are thankful to P. Grigoriev, K. Kanoda, S. Khasanov, M. Lang, A. Pustogow, R. Ramazashvili, R. Valenti, S. Winter, and V. Zverev for stimulating and illuminating discussions. We are also grateful to M. Lang for providing a test sample of $\kappa$-Cl. The work was supported in part by the German Research Foundation (Deutsche Forschungsgemeinschaft, DFG) via Grant No. KA 1652/5-1 and Grant No. GR 1132/19-1.
N.D.K. acknowledges the partial support from the RFBR grant No. 21-52-12027.


\begin{thebibliography}{96}
\bibitem{comm_k-Br}{Even more close to $\kappa$-Cl is the ambient-pressure superconductor $\kappa$-(BEDT-TTF)$_2$Cu[N(CN)$_2$]Br. 
   However, in this salt the $\beta$ oscillations have only been seen at $B>17$\,T, well beyond our field range \cite{weis97,weis99a}}
    \BibitemShut   {NoStop}%
\bibitem{comm_PhysChemPres}{The analogy between the physical pressure and 
  ``chemical pressure'' should be taken with care. For example, a comparison 
  between the $\kappa$-Cl and $\kappa$-NCS salts in 
  Sec. III\,B of the main text, shows that it may break in the case of anion substitution}
  \BibitemShut {NoStop}%
\end{thebibliography}

\newpage
\begin{center}
{\large {\bf Supplementary Material for:\\
Coherent heavy charge carriers in an organic conductor near the bandwidth-controlled Mott transition}}\\
by S. Oberbauer, S. Erkenov, W. Biberacher, N. D. Kushch, R. Gross, and M.~V.~Kartsovnik
\end{center}

\setcounter{section}{0}
\makeatletter
\makeatother

\setcounter{figure}{0}
\makeatletter
\makeatother

\makeatletter
\renewcommand \thesection{S-\@Roman\c@section}
\renewcommand \thefigure{S\@arabic\c@figure}
\renewcommand \theequation{S\@arabic\c@equation}

\makeatother



\section{Experimental details}

\begin{figure}[b]
\center
\includegraphics[width = 0.75 \columnwidth]{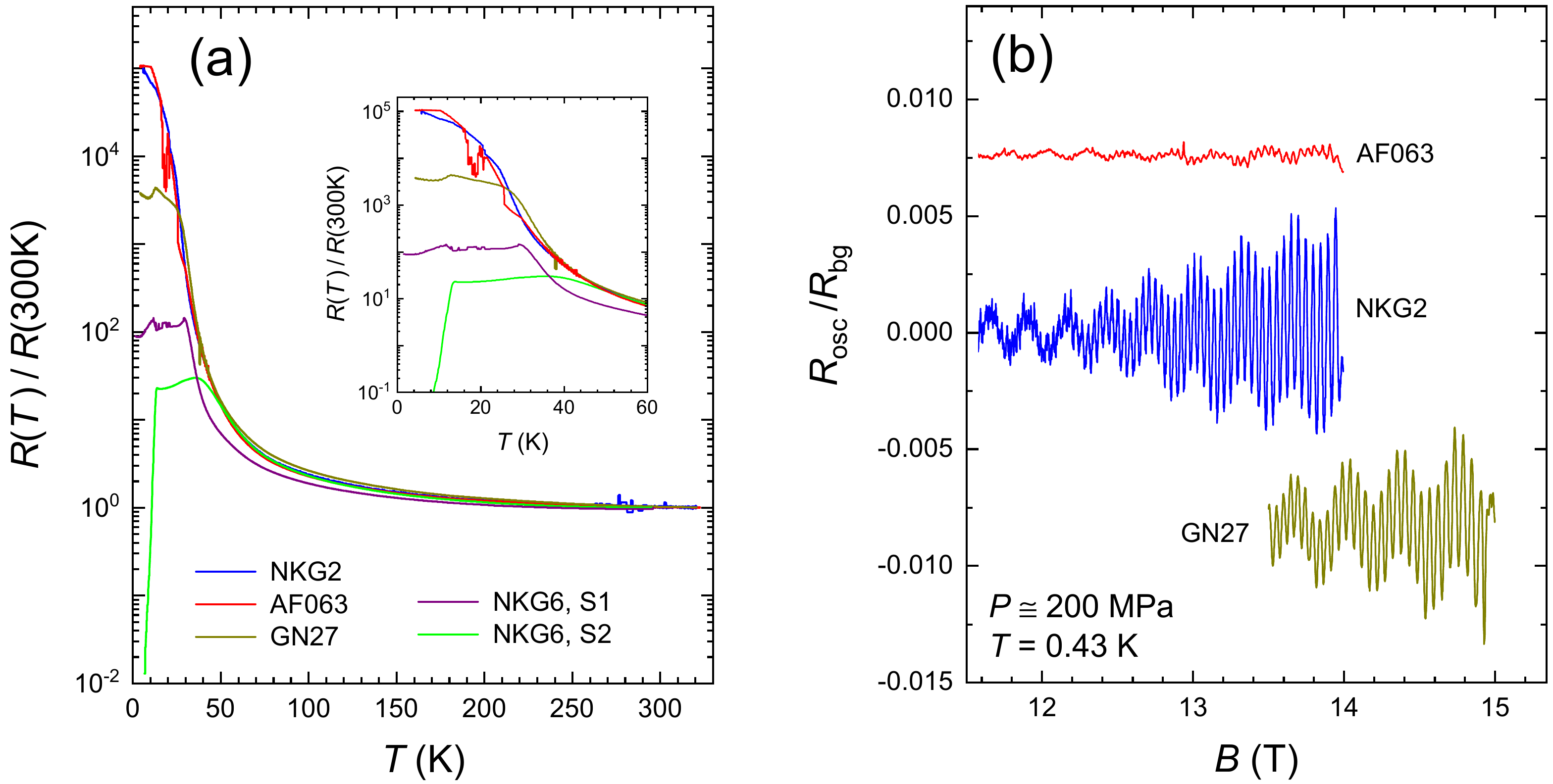}
\caption{(a) Examples of the zero-pressure temperature dependence of the interlayer resistance of $\kappa$-Cl samples tested for the experiment. Inset: a close-up of the data at low temperatures. (b) SdH oscillations in three samples selected for testing under pressure, in magnetic fields. All three samples display oscillations with two fundamental frequencies $F_{\alpha}$ and $F_{\beta}$, however, with different amplitudes. }
\label{Rtests}
\end{figure}

\textit{Sample selection -- }For a detailed examination of the direct neighborhood of the Mott state in $\kappa$-(BEDT-TTF)$_2$Cu[N(CN)$_2$]Cl ($\kappa$-Cl) in the  low-$P$ range, $20\mathrm{\,MPa} < P < 100\mathrm{\,MPa}$, particular care should be taken about the sample quality and homogeneity. To this end, we have tested several crystals from different batches.
As a first step, zero-pressure resistance was measured to filter out crystals with a significant residual conductance and especially with traces of the superconducting transition in the low-temperature Mott-insulating state indicating crystal disorder and/or significant internal strain. Some zero-pressure $R(T)$ examples
are displayed in Fig.\,\ref{Rtests}(a).

Samples "NKG2", "AF063", and "GN27" showed the best $R(T)$ characteristics and were further tested for magnetoresistance under a pressure of $\sim 200$\,MPa at liquid helium-3 temperatures. All three samples showed Shubnikov-de Haas (SdH) oscillations,
however with different amplitudes. For the detailed low-$P$ SdH experiment in the dilution fridge sample "NKG2" (labelled as ``sample \# 1'' in the main text) exhibiting the strongest oscillations has been selected. Additionally, sample ``GN27'' (``sample \# 2'' in the main text) was used for measurements in an extended pressure range at $^3$He temperatures, $0.43 \leq T \leq 1.1$\,K.

\begin{figure}[t]
\center
\includegraphics[width = 1.0 \columnwidth]{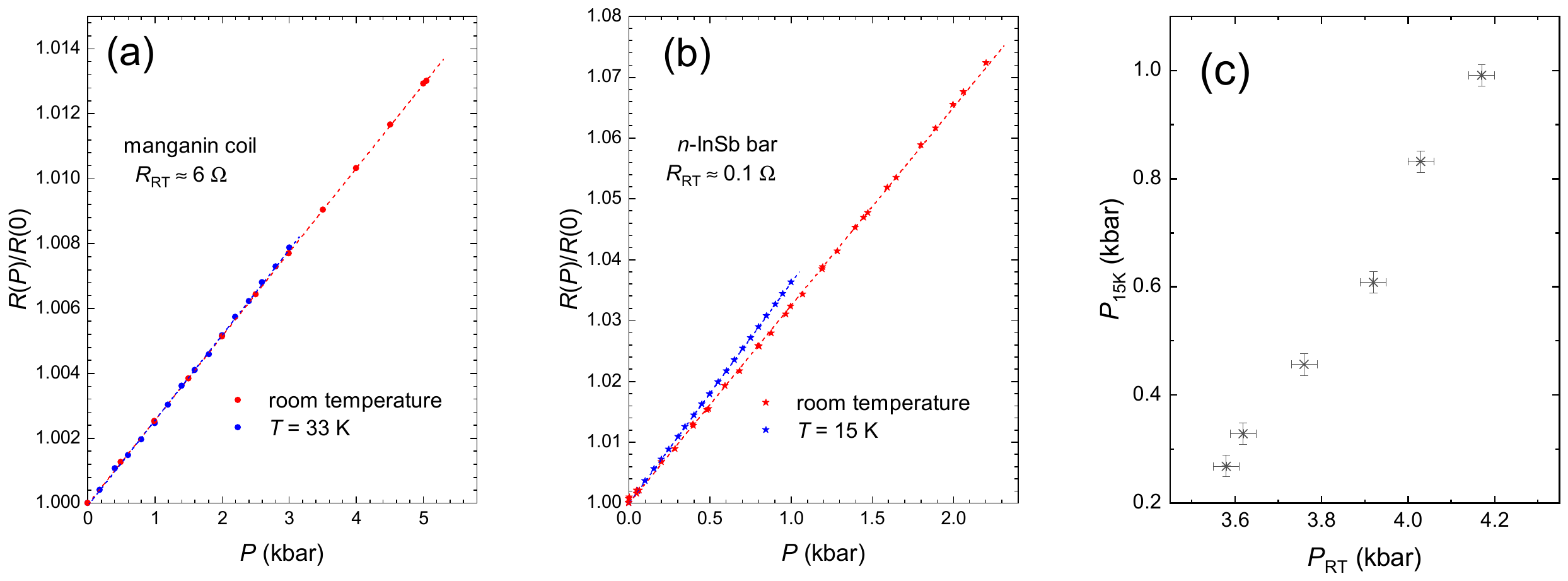}
\caption{(a) Pressure dependence of the normalized resistance of a thin manganin wire trained and calibrated in a He-gas pressure setup using a commercial membrane pressure gauge. The data obtained at room temperature (RT) and at a low temperature are shown. Dashed lines are linear fits yielding the slope $0.259 \%/{\mathrm{kbar}}$ and $0.268 \%/{\mathrm{kbar}}$ at RT and 33\,K, respectively. (b) Same for a $n$-doped InSb bar cut out of a commercially available pressure gauge SPG10 (Institute of High Pressure Physics Unipress, Warsaw, Poland). Linear fits (dashed lines) show the slopes $3.25 \%/{\mathrm{kbar}}$ and $3.63 \%/{\mathrm{kbar}}$ at RT and 15\,K, respectively. (c) Low-temperature pressure in the cell, determined at 15\,K, plotted against the pressure applied at RT.}
\label{Psens}
\end{figure}

\textit{Pressure control -- }Besides the high sample quality, precise pressure control was crucial for our study. Our cryomagnetic setup
restricts the choice of high-pressure technique to a small ($\leq 4$\,mm diameter of the sample channel) piston-cylinder clamp pressure cell. The cell used in the experiment was made of nonmagnetic beryllium bronze purified to minimize paramagnetic impurities. The pressure in the sample channel was generated using silicone oil GKZh as a pressure medium and monitored via a manganin or a $n$-doped InSb resistive sensor. Both sensors exhibit a monotonic, close to linear pressure dependence that only slowly varies with temperature. They were calibrated in a He-gas pressure setup at room temperature and at low temperatures, revealing a sensitivity of $0.25\%/\mathrm{kbar}$ and $\simeq 3.5\%/\mathrm{kbar}$ ($1\mathrm{\,kbar} = 100\mathrm{\,MPa}$) for the manganin and InSb sensors, respectively, in the pressure range of interest, see Fig.\,\ref{Psens}(a),(b). For the measurements, the sensor was mounted into the cell very close to the samples, see Fig.\,2(a) of the main text.
As one can see in Fig.\,\ref{Psens}(c), the pressure in our setup significantly decreases upon cooling from room temperature. All the pressure values given in the text were determined at $T = 15$\,K.

The measurements in the range 40\,-- 300\,MPa on sample \#2 were done with the manganin sensor, allowing to determine pressure with an accuracy of $\pm 10$\,MPa. For the more precise low-$P$ measurements on sample \#1 the InSb sensor was used, reducing the pressure error bar to $\pm 2$\,MPa (that is within the size of the symbol in Fig.\,3 of the main text).

\section{Estimation of the magnetic-breakdown gap in $\kappa$-Cl}

The quantum oscillations with the frequency $F_{\beta} \simeq 4\times10^3$\,T ($\beta$ oscillations) in the $\kappa$-salts are caused by magnetic breakdown (MB) through four identical small gaps separating the closed portion $\alpha$ of the Fermi surface from the open sheets extended along the $k_x$-direction, see inset in Fig.\,1 of the main text. In moderately strong magnetic fields both the $\alpha$ and $\beta$ oscillations can be observed, with the amplitudes depending on the gap value $\Delta_{\mathrm{MB}}$ through the MB damping factors \cite{wosn96,kart04}:
\begin{align}
  R_{\mathrm{MB},\alpha} = \left[ 1 - \exp\left({-\frac{B_{\mathrm{MB}}}{B}} \right) \right]\,,\label{RMBa}\\
  R_{\mathrm{MB},\beta} = \exp\left({-\frac{2B_{\mathrm{MB}}}{B}} \right),\label{RMB}
\end{align}
respectively, with the characteristic MB field \cite{shoe84}
\begin{equation}
B_{\mathrm{MB}} \simeq \frac{m_c}{\hbar e}\frac{\Delta_{\mathrm{MB}}^2}{\varepsilon_{\mathrm{F}}}\,,
\label{BMB}
\end{equation}
$\varepsilon_{\mathrm{F}}$ being the Fermi energy and $m_c$ the cyclotron frequency.

The MB factors should be incorporated into the standard Lifshitz-Kosevich formula \cite{shoe84} for the corresponding oscillation amplitudes:
\begin{equation}
A_{i} \equiv \frac{\sigma_{\mathrm{osc},i}}{\sigma_{0,i}} \propto \sqrt{B}m_{c,i}R_{T,i}R_{\mathrm{D},i},R_{\mathrm{MB},i},
\label{LKMB}
\end{equation}
where $\sigma_{\mathrm{osc},i}$ and $\sigma_{0,i}$  are, respectively, the oscillating and classical (nonoscillating) components of the conductivity provided by the carriers on the $i$-th orbit of the Fermi surface, $m_{c,i}$ the relevant cyclotron mass, and
\begin{equation}\label{RT}
R_{T,i} = \frac{Km_{c,i}T/B}{\sinh(Km_{c,i}T/B)}
\end{equation}
and
\begin{equation}\label{RD}
R_{\mathrm{D},i} = \exp(-Km_{c,i}T_{\mathrm{D}}/B)
\end{equation}
the corresponding temperature and Dingle damping factors, respectively, with $K=2\pi^2k_{\mathrm{B}}/\hbar e$.
Given all the other parameters in Eqs.\,(\ref{LKMB})-(\ref{RD}) for the Fermi surface orbits $\alpha$ and $\beta$ are known, we can estimate $B_{\mathrm{MB}}$ from the amplitude ratio $A_{\alpha}/A_{\beta}$.

We use the Shubnikov-de Haas (SdH) data obtained for $\kappa$-Cl at temperature $T = 100$\,mK, under pressure $P \approx 99$\,MPa as shown in Fig.\,3 of the main text. At this pressure both the $\alpha$ and $\beta$ oscillations are clearly resolved in the field range between 12 and 15 T and the ratio between the corresponding FFT amplitudes is $A_{\alpha}/A_{\beta} \approx 0.13$.
For our rough estimation we assume the classical conductivity ratio  $\sigma_{0,\alpha}/\sigma_{0,\beta} = 2$. Indeed, the $\alpha$ pocket accommodates approximately one half of the electronic states on the Fermi surface and therefore provides roughly one half of the total interlayer conductivity: $\sigma_{0,\alpha} \simeq \sigma_{0}/2$, whereas the $\beta$ orbit encompasses the entire Fermi surface, i.e. $\sigma_{0,\beta} = \sigma_0$.
With this assumption and plugging the experimental values  $m_{c,\alpha} \approx 2.5 m_0$, $m_{c,\beta} \approx 5.2 m_0$, $T_{\mathrm{D}} \simeq 0.5$\,K, and $1/B = \left(\frac{1}{12\mathrm{\,T}}+\frac{1}{15\mathrm{\,T}}\right)/2\ = 0.075$\,T$^{-1}$ (the midpoint of the relevant inverse-field window) in Eqs.\,(\ref{LKMB})-(\ref{RD}), we obtain the MB field, $B_{\mathrm{MB}} \simeq 1.3$\,T.

Now we use Eq.\,(\ref{BMB}), which defines $B_{\mathrm{MB}}$ through the Fermi energy and the MB gap $\Delta_{\mathrm{MB}}$, to estimate the latter. To this end, we take into account an almost circular shape of the $\beta$ orbit \cite{yama96}
so that
\begin{equation}\label{EF}
\varepsilon_{\mathrm{F}} \sim \frac{\hbar^2k_{\mathrm{F}}^2}{2m_{c,\beta}} \simeq \frac{\hbar e}{m_{c,\beta}}F_{\beta}\,,
\end{equation}
where we made use of the Onsager relation between the SdH frequency $F$ and the relevant Fermi-surface cross-section area $S_{\mathrm{FS}}$: $F = \frac{\hbar}{2\pi e}S_{\mathrm{FS}} \simeq \frac{\hbar k_{\mathrm{F}}^2}{2e}$.
A substitution of Eq.\,(\ref{EF}) into Eq.\,(\ref{BMB}) with $F_{\beta} = 3850$\,T and $B_{\mathrm{MB}} = 1.3$\,T, yields the MB gap,
\begin{equation}
 \Delta_{\mathrm{MB}} \sim \frac{\hbar e}{m_{c,\beta}} \sqrt{F_{\beta}B_{\mathrm{MB}}} \approx 1.6\,\mathrm{meV}\,.
\end{equation}
The same evaluation made for $P = 46$\,MPa with $A_{\alpha}/A_{\beta} \approx 0.083$, $m_{c,\beta} \approx 2m_{c,\alpha} \approx 5.9 m_0$, and $T_{\mathrm{D}} \approx 0.9$\,K yields $B_{\mathrm{MB}} \simeq 0.9$\,T and $\Delta_{\mathrm{MB}} \sim 1.2$\,meV.

\section{Cyclotron mass determination for $\kappa$-Cl}

\begin{figure}[ht!]
\center
\includegraphics[width = 0.7 \columnwidth]{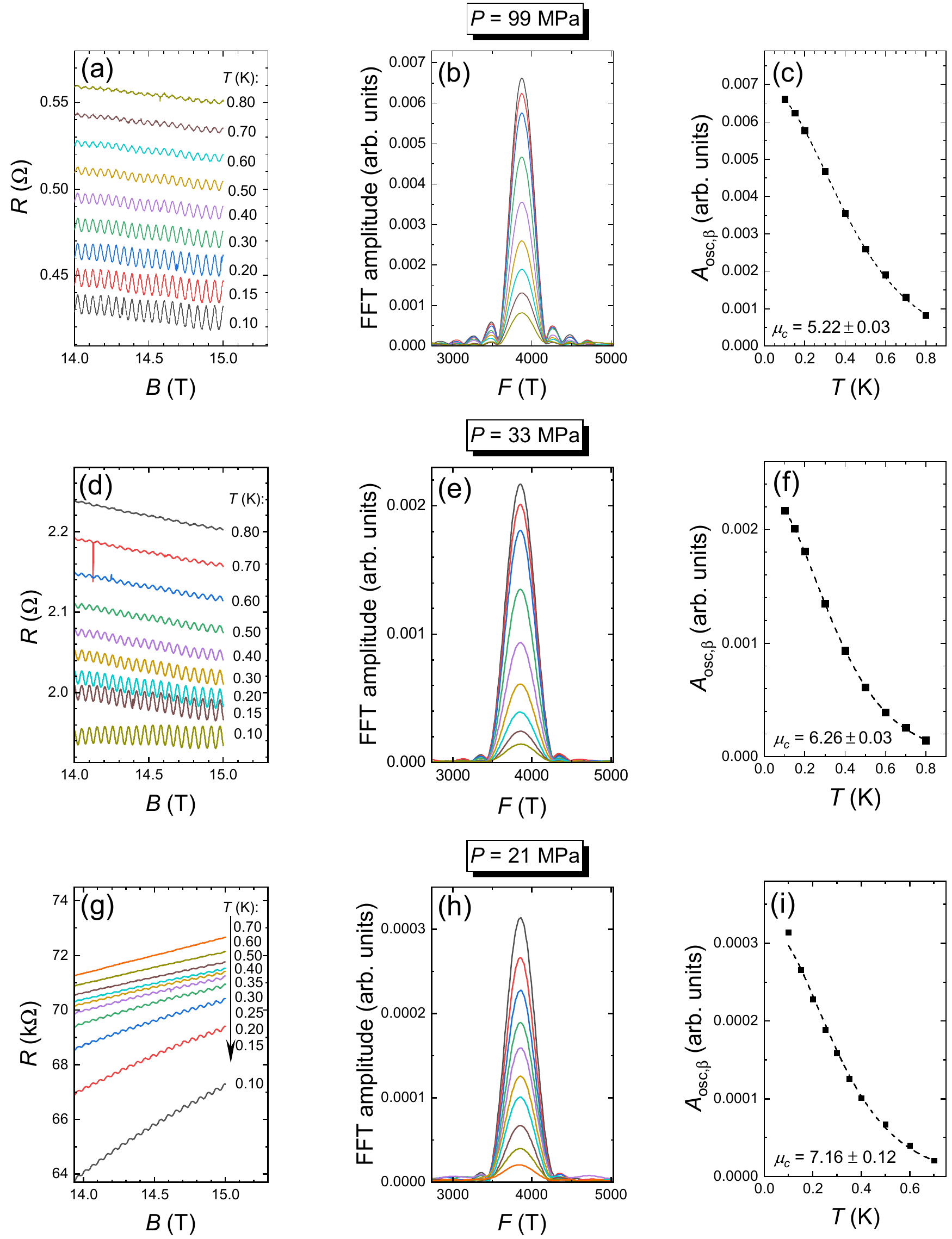}
\caption{(a) High-field fragments of the $B$-dependent interlayer resistance of $\kappa$-Cl sample \#1 recorded at different temperatures, at $P=99$\,MPa.
(b) The fast Fourier transform (FFT) of the oscillatory component of the signal in (a), normalized by the nonoscillating background resistance, exhibits a peak at $F_{\beta} \approx 3850$\,T.
The same colour code as in (a) is used for different temperatures. (c) The height of the FFT peak as a function of temperature. The fit using Eq.\,(\ref{RT}) (dashed line) yields the effective cyclotron mass $\mu_c \equiv m_{c,\beta}/m_0 = 5.22$ ($m_0$ is the free electron mass).
Panels (d)--(f) and (g)--(i) show the analogous data for $P = 33$\,MPa and 21\,MPa, respectively. The curves in (a) for $T \geq 0.15$\,K are shifted vertically for clarity.}
\label{mass1}
\end{figure}

\begin{figure}[tb]
\center
\includegraphics[width = 0.7 \columnwidth]{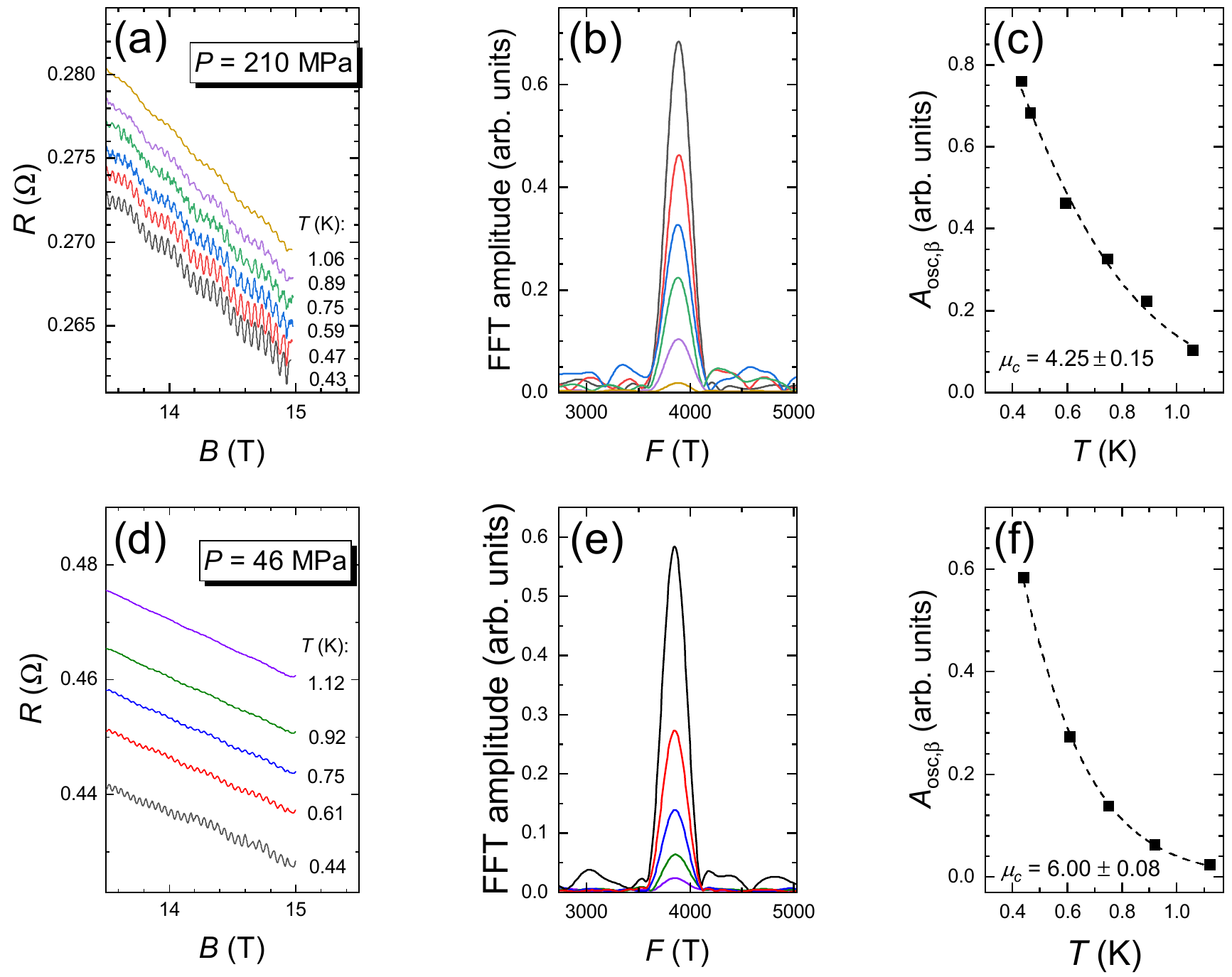}
\caption{Same plots as in Fig.\,\ref{mass1} but for sample \#2 at: (a)--(c) $P = 209$\,MPa and (d)--(f) $P=46$\,MPa.}
\label{mass2}
\end{figure}

The effective cyclotron mass was evaluated by fitting the $T$-dependence of the SdH amplitude with the Lifshitz-Kosevich temperature damping factor $R_T$, given by Eq.\,(\ref{RT}).
The amplitude $A_{\mathrm{osc}}(T)$ was determined from the FFT spectrum of the normalized oscillatory signal $\rho_{\mathrm{osc}}/\rho_{\mathrm{bg}}$, where $\rho_{\mathrm{bg}}$ is the nonoscillating (background) field- and temperature-dependent resistivity. The field windows for the FFT analysis, $14\mathrm{\,T} \leq B \leq 15$\,T for sample \#1 and $13.5\mathrm{\,T} \leq B \leq 15$\,T for sample \#2, were chosen narrow enough so as to avoid a spurious effect of the $B$-dependence of the SdH signal on the FFT amplitude \cite{audo18}, at the same time keeping a sufficient number of oscillation periods within the window.
Figs.\,\ref{mass1} and \ref{mass2} show a few examples illustrating the evaluation procedure. In addition to the above-mentioned samples, two other samples were checked at a few pressures, confirming the presented results within the experimental error bar.

A comment should be made on a potential influence of superconductivity on the oscillation behavior in the phase coexistence region. Indeed, inside this region
superconductivity becomes considerably more robust against magnetic field, persisting up to the highest fields applied at the lowest temperatures, see
$R(B)$ curves corresponding to $P=33,27,$ and 21\,MPa in Fig.\,2 of the main text. The nucleation of superconductivity can also be traced in Fig.\ref{mass1}(d),(g), where even at the highest field the nonoscillating resistance background drops with an increasing rate upon cooling below 0.15\,K and 0.35\,K at 33 and 21\,MPa, respectively. In general, the superconducting onset introduces an additional $T$- and $B$-dependent damping mechanism for quantum oscillations \cite{mani01}.
While this effect is pronounced in a number of conventional type-II superconductors, it seems to be weaker in materials in which superconductivity occurs at the border of a competing ordering ground state, see, e.g., Ref.\,\citenum{hsu21} and references therein. Therefore, taking into account that in our case the resistance drop caused by superconductivity is small ($< 10\%$, in the field range of interest) we do not expect it to have a significant effect on the SdH signal. In particular, we do not observe an increase of the Dingle temperature that could be attributed to an additional damping of the oscillations in the mixed state. Moreover, we note that any noticeable change of the oscillation amplitude in the mixed state would only lead to an underestimation of the effective mass at low pressures, i.e. act in the direction opposite to the observed steepening of the $m_c(P)$ dependence in the phase-coexistence region.

\section{Quantum oscillations in $\kappa$-NCS}
\begin{figure}[tb]
\center
\includegraphics[width = 0.7 \columnwidth]{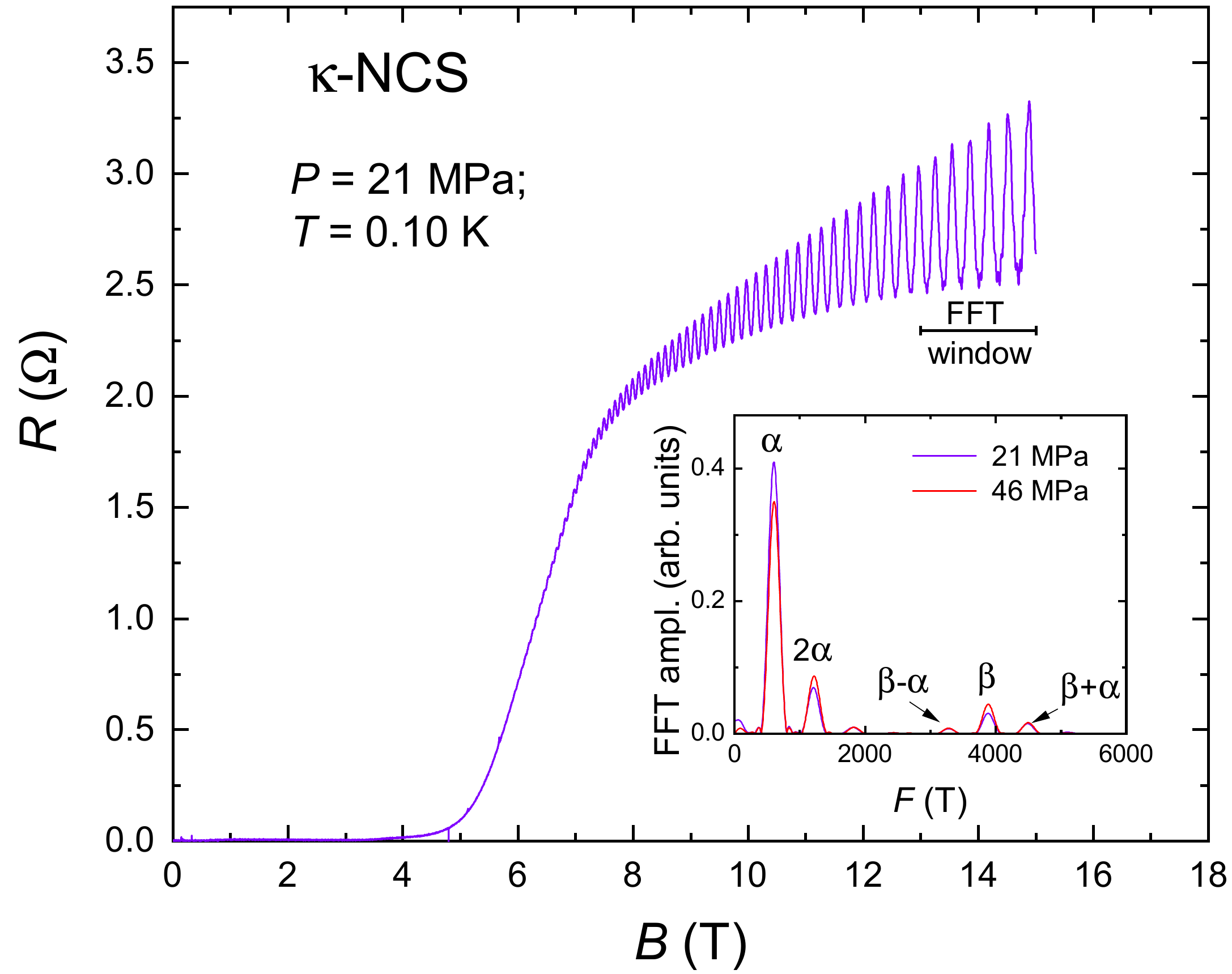}
\caption{Example of the $B$-dependent interlayer resistance of the $\kappa$-NCS sample at the lowest pressure, $P=21$\,MPa. Inset: FFT spectrum of the SdH oscillations at this pressure as well as at $P=99$\,MPa. The FFT was made in the field window from 13 to 15\,T (indicated by the horizontal bar in the main panel). The spectrum is dominated by the oscillations associated with the classical $\alpha$ orbit on the Fermi surface and its second harmonic. Additionally, the MB oscillations $\beta$ and the combination frequencies are clearly resolved.}
\label{MR-NCS}
\end{figure}
\begin{figure}[tb]
\center
\includegraphics[width = 0.7 \columnwidth]{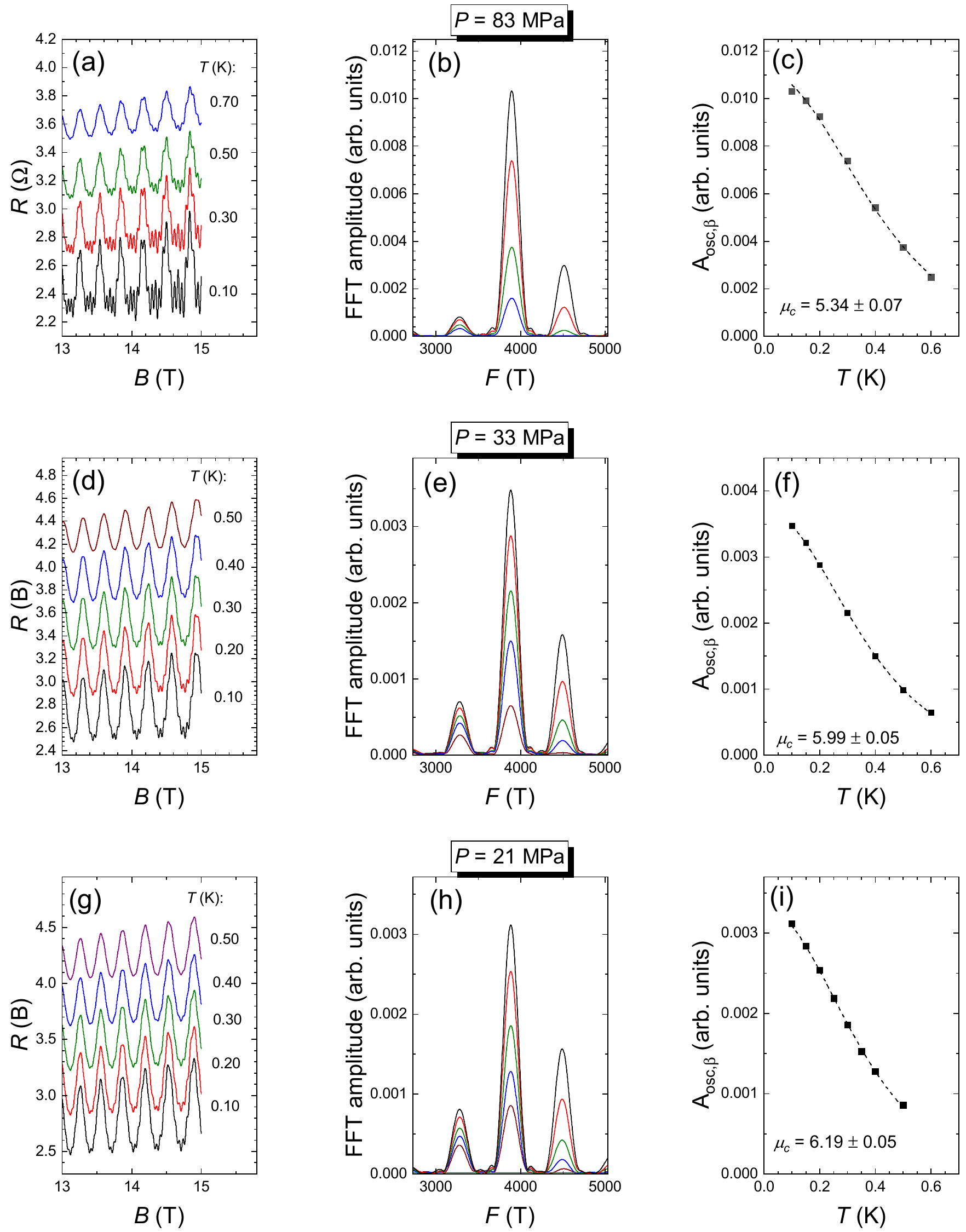}
\caption{(a),(d),(g) -- High-field fragments of the $B$-dependent interlayer resistance of the $\kappa$-NCS sample recorded at three different pressures, at different temperatures; the curves are vertically shifted for clarity. (b),(e), and (g) -- The corresponding FFT fragments around the frequency $F_{\beta} \approx 3850$\,T. The color code is the same as for the $R(B)$ curves. (c),(f),(i) -- $T$-dependence of the $\beta$-oscillation amplitude fitted with Eq.\,(\ref{RT}). The resulting values of the effective cyclotron mass are shown in the graphs in units of the free electron mass.
}
\label{massNCS}
\end{figure}
As noted in the main text, $\kappa$-(BEDT-TTF)$_2$Cu(NCS)$_2$ ($\kappa$-NCS) is
one of the closest analogues of $\kappa$-Cl but metallic already at ambient pressure
which makes it a perfect reference material for our purposes \cite{comm_k-Br}. The experimental data on the $\beta$ oscillations under pressure are extremely scarce, most likely due to the relatively high MB field, $B_{\mathrm{MB}} \sim 15-30$\,T \cite{sasa91a,meye95,audo16}. We are currently aware of just a single publication \cite{caul94} reporting on the SdH $\beta$ oscillations at three pressures, all above 1\,GPa, that is, much higher than the pressure range of our interest. Therefore we have performed a purposeful study of the $\beta$ oscillations
in $\kappa$-NCS at $P < 100$\,MPa, at temperatures down to 0.1\,K. To facilitate a direct comparison, we have measured the oscillations in parallel with the $\kappa$-Cl sample \#1 placed next to the $\kappa$-NCS sample, see Fig.\,2(a) of the main text.

Figure \ref{MR-NCS} shows an example of the field-dependent interlayer resistance typical of a high-quality $\kappa$-NCS crystal. SdH oscillations can already be resolved in the superconducting mixed state, starting from 6\,T. The dominant SdH signal with the frequency $F_{\alpha} = 606$\,T comes from the classical orbit $\alpha$ on the Fermi surface. The amplitude of the MB $\beta$ oscillation is $\simeq 15$ times smaller even at the highest fields, $13-15$\,T at $P=21$\,MPa, -- this is seen from the FFT data in the inset in Figure \ref{MR-NCS}. At increasing pressure the relative amplitude of the $\beta$ oscillations moderately increases (see the FFT spectrum for $P=46$\,MPa in the inset), which suggests a reduction of the MB gap. The frequency of the $\beta$ oscillations, $F_{\beta} = 3890$\,T, differs from that in $\kappa$-Cl by less than $1\%$. This is not surprising: the area enclosed by the $\beta$ orbit is equal to the 1st Brillouin zone area and the latter is almost the same in the two salts, according the crystal structure data \cite{geis91,schu91}. We note that in contrast to the $\beta$ frequency, the $\alpha$ frequency shows a much larger difference, $\approx 10\%$. This indicates significantly different inplane anisotropies of the Fermi surfaces in $\kappa$-Cl and $\kappa$-NCS.

Besides the fundamental frequencies $F_{\alpha}$ and $F_{\beta}$, the FFT reveals a very strong 2nd harmonic $2F_{\alpha}$ and the combination frequencies $F_{\beta \pm \alpha}$. These frequencies are known to originate from the highly two-dimensional character of the electronic system of $\kappa$-NCS, see, e.g., \cite{kart04}. The two-dimensionality may lead to significant deviations of the oscillation behavior from the standard Lifshitz-Kosevich theory as it concerns the field-dependent SdH spectrum \cite{kart04,grig03,grig12}. However the temperature dependence of the fundamental-harmonic amplitude is still adequately described by the temperature damping factor given by Eq.\,(\ref{RT}), at least as long as the amplitude is small. At our conditions, the amplitude of the $\beta$ oscillations never exceeds $10\%$ of the background resistivity, which justifies the use of Eq.\,(\ref{RT}) for evaluation of the effective cyclotron mass.

Examples of the oscillating magnetoresistance and the corresponding mass plots are shown in Fig.\,\ref{massNCS} for the highest pressure as well as for the two lowest pressures. One can see that even at the lowest pressures, where the relative amplitude of the $\beta$ oscillations is the weakest, we have sufficient data for evaluating the mass to an accuracy of $\simeq 1\%$. The results are presented in Fig.\,4 of the main text.

\section{Effective mass in other bandwidth-controlled Mott materials}

While an enhancement of the effective mass upon approaching the MIT is commonly considered as a well established phenomenon, explicit experimental data, which could be used for a quantitative comparison with theoretical predictions, is extremely scarce.
Very recently, an increase of the effective mass and a simultaneous increase of the correlation strength ratio $u \equiv U/W$ \cite{Comm_W-t}
has been reported, based on infrared conductivity studies, for two other organic salts, $\kappa$-[(BEDT-STF)$_x$(BEDT-TTF)$_{1-x}$]$_2$Cu$_2$(CN)$_3$ \cite{pust21}, and $\beta'$-EtMe$_3$Sb[Pd(dmit)$_2$]$_2$ \cite{li19}. The mass renormalization was evaluated by applying the extended Drude (ED) model or from the spectral weight (SW) analysis, see Ref. \citenum{li19} for details. The $U$ and $W$ values have also been estimated by multiple-component fitting of the optical conductivity spectra.

\begin{figure}[tb]
\center
\includegraphics[width = 1.0 \columnwidth]{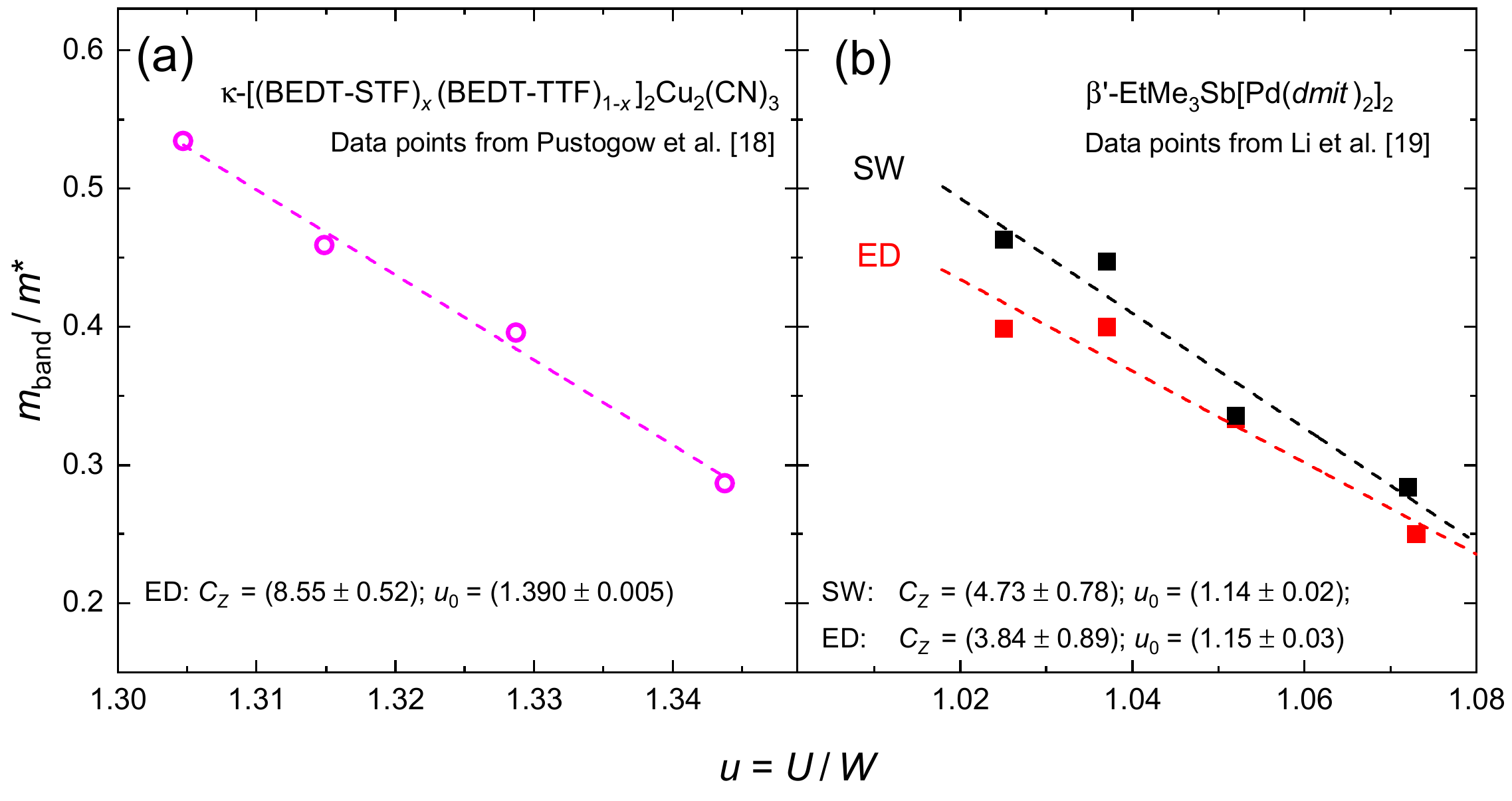}
\caption{Dependence of the inverse mass-renormalization factor $m_{\mathrm{band}}/m^{\ast}$ on the correlation strength ratio $u \equiv U/W$ obtained from infrared conductivity measurements on the organic charge-transfer salts: (a) $\kappa$-[(BEDT-STF)$_x$(BEDT-TTF)$_{1-x}$]$_2$Cu$_2$(CN)$_3$; data taken from Ref. \citenum{pust21}; and (b) $\beta'$-EtMe$_3$Sb[Pd(dmit)$_2$]$_2$; data taken from Ref. \cite{li19}. The mass renormalization factor was evaluated by applying the extended Drude (ED) or spectral-weight (SW) analysis.}
\label{m-optic}
\end{figure}

For $\kappa$-[(BEDT-STF)$_x$(BEDT-TTF)$_{1-x}$]$_2$Cu$_2$(CN)$_3$, $u$ was varied by means of changing $x$: the $x=0$ salt is insulating; the compositions with $x \gtrsim 0.12$ are metallic \cite{pust21}. A comparison of the $T$--$P$ and $T$--$x$ phase diagrams of $\kappa$-[(BEDT-STF)$_x$(BEDT-TTF)$_{1-x}$]$_2$Cu$_2$(CN)$_3$ suggests that the effect of chemical substitution is similar to pressure \cite{pust21a,sait21}: an increase of $x$ by 0.1 corresponds to an increase of pressure by roughly $2$\,kbar \cite{comm_PhysChemPres}.
In Fig.\,\ref{m-optic}(a) we plot the values of the inverse renormalization factor $m_{\mathrm{band}}/m^{\ast}$, given in the inset to Fig.\,3(h) of Ref. \citenum{pust21} for different $x$, against the corresponding $u$ values taken from the Supplementary Figure 6 of that work. Similarly to our data in Fig.\,4 of the main text, the data in Fig.\,\ref{m-optic}(a) can be nicely fitted by the linear function,
\begin{equation}\label{BR2}
\frac{m_{\text{band}}}{m^{\ast}} = C_Z\left(1-\frac{u}{u_0}\right),
\end{equation}
which is equivalent to Eq.\,(2) of the main text. Moreover, similarly to our case, the coefficient $C_Z \approx 8.5$ is an order of magnitude higher than the value expected from DMFT \cite{geor96}.

In Fig.\,\ref{m-optic}(b), we present the $u$-dependence of the inverse mass renormalization factor for $\beta'$-EtMe$_3$Sb[Pd(dmit)$_2$]$_2$, using the data reported in Fig.\,4(d),(e) of Ref. \citenum{li19}; here, the electronic state was tuned by changing pressure. Despite the relatively large scattering and some model dependence of the evaluation, the data is apparently consistent with Eq.\ref{BR2}. The slope is a factor of $\sim 2$ smaller than in Fig.\,\ref{m-optic}(a), however, still significantly exceeds the theoretical predictions.

As compared to the quantum oscillation analysis, the mass evaluation from the optical data may depend on the specific model applied (this already can be noticed from the difference between the data in Fig.\,\ref{m-optic}(b) based on the SW and ED models, respectively). Another point to keep in mind is that the optical experiments are essentially sensitive to the properties near the surface of a metal, which may differ, especially in the phase-coexistence region, from the bulk properties probed by the quantum oscillation technique.
It would be highly interesting to perform a quantum-oscillation study of the abovementioned two compounds for a direct comparison with our $\kappa$ salts. Unfortunately, the inherent disorder in the conducting layers of the mixed salt $\kappa$-[(BEDT-STF)$_x$(BEDT-TTF)$_{1-x}$]$_2$Cu$_2$(CN)$_3$ makes the observation of the oscillations in realistic magnetic fields $<100$\,T highly unlikely. In $\beta'$-EtMe$_3$Sb[Pd(dmit)$_2$]$_2$, the disorder potential is mainly localized in the insulating cation layers \cite{kato22}, which may be less critical for the electron scattering in the conducting anion layers and hence more promising for a quantum oscillation experiment.


As for inorganic bandwidth-controlled Mott insulators, one of the most extensively studied is vanadium sesquioxide V$_2$O$_3$, see, e.g., \cite{imad98,cart93,hans13,deng14,mcle17}
and references therein. Like in our salts, the ground state of this material can be conveniently tuned by applying pressure of the order of 1\,GPa \cite{imad98,cart93}. Heat capacity measurements on lightly doped V$_{1.987}$O$_3$ \cite{cart93} under pressure have revealed a strong increase of the Sommerfeld coefficient $\gamma \propto m^{\ast}$ at approaching the MIT from the metallic side. The increase was attributed to the enhancement of the electron effective mass renormalization in accord with the Brinkmann-Rice model \cite{cart93}. In the interval between $0.57$ and 0.25\,GPa, $\gamma$ changed by a factor of $\approx 2.3$, see Fig.\,5 of Ref.\,\citenum{cart93}. To make a rough comparison with our results,
an increase of pressure by 100 MPa from the critical pressure of the MIT ($P_{\mathrm{MIT}}\approx 230$\,MPa) led to the relative change of the mass, $\frac{\Delta m^{\ast}}{m^{\ast}} = \frac{\Delta\gamma}{\gamma} \approx 0.3$. This is even stronger than the respective change in our salts $\frac{\Delta m^{\ast}}{m^{\ast}} \approx 0.2$ in the pressure interval 0 to 100\,MPa (Fig.\,4 of the main text).
Taking into account that compressibility of V$_2$O$_3$, $\beta \equiv \frac{1}{V}\frac{dV}{dP} = 0.67\times 10^{-2}$\,GPa$^{-1}$ \cite{ovsy13} is an order of magnitude lower than in the relatively soft organic $\kappa$-salts ($\beta \approx 6.5 \times 10^{-2}$\,GPa$^{-1}$ \cite{schu94,raha97}), it is very likely that the effective mass  here is even more sensitive to the correlation strength ratio than in our materials. However an explicit quantitative comparison between the experiment and theory is still to be done.

We note that the theoretical analysis in the case of V$_2$O$_3$ is complicated by the presence of an incommensurate antiferromagnetic order in the metallic phase (contributing to the heat capacity), structural instability, and deviations from the exact half-filling. In this respect, the organic $\kappa$ salts provide a much simpler
and convenient platform for confrontation of the theory and experiment.

\section{Evaluation of $T_{\mathrm{D}}$}

\begin{figure}[ht!]
\center
\includegraphics[width = 0.85 \columnwidth]{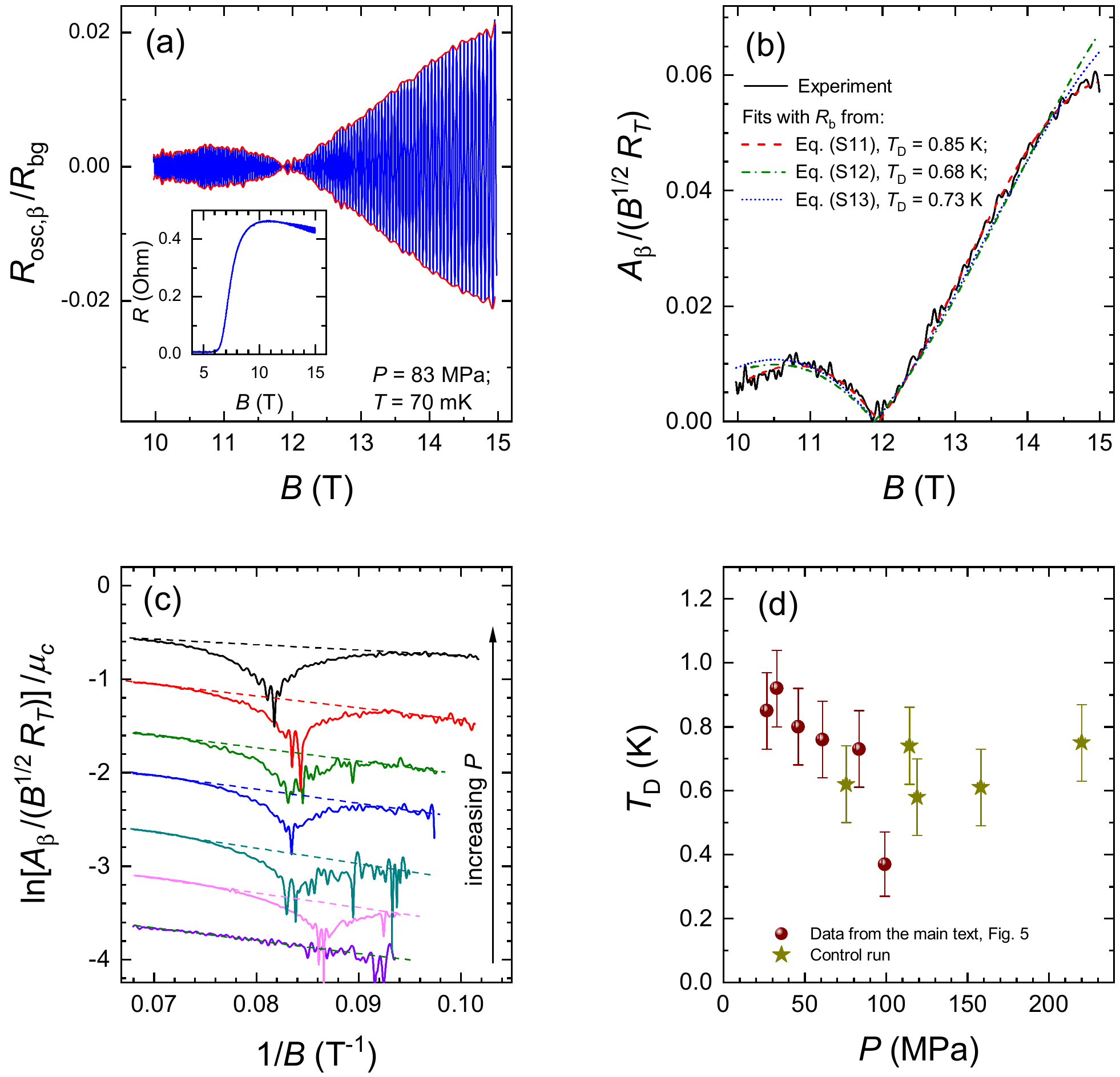}
\caption{(a) $\beta$ component of the SdH oscillations in $\kappa$-Cl sample $\# 1$ at $P = 83$\,MPa, $T = 70$\,mK. The red lines are envelopes of the oscillating signal used for determination of the oscillation amplitude $A_{\beta}(B)$ used for the plots in panels (b) and (c). The data is obtained by FFT-filtering of the high-field fragment of the $B$-dependent resistance shown in the Inset.
(b) Ratio $\frac{A_{\beta}}{B^{1/2}R_T}$ vs. $B$ (solid black line) fitted according to Eq.\,(\ref{ABeats}), using for $R_{\mathrm{b}}$ the expressions given by Eq.\,(\ref{Rb1}) (dashed red line), Eq.\,(\ref{Rb-18}) (dash-dotted green line), and Eq.\,(\ref{Rb3}) (dotted blue line). The respective $T_{\mathrm{D}}$ values obtained from the fits are indicated in the graph.
(c) Dingle plots of the SdH amplitude (see text) at $T = 70 \pm 5$\,mK,  for pressures [MPa]:
21, 27, 33, 46, 61, 83, and 99.
The curves are vertically shifted for clarity. The dashed straight lines are tangents to the Dingle plots.
(d) Dingle temperature values obtained in the main experiment (circles) along with those obtained in the higher-pressure control run at $^3$He temperatures (stars). }
\label{TD-plots}
\end{figure}

For estimating the Dingle temperature, the field dependence of the oscillation amplitude was studied in an extended range, 10 to 15\,T, at $T= 70 \pm 5$\,mK. Fig.\,\ref{TD-plots}(a) shows an example of the $\beta$ oscillations at $P = 83$\,MPa; the raw resistance $R(B)$ is plotted in the inset. The oscillation amplitude $A$ was determined as the difference between the upper and lower envelopes (red lines in Fig.\,\ref{TD-plots}(a)).

The oscillation pattern exhibits prominent beating, which most likely originates from the equal contributions of two slightly different extremal cross-sections of the weakly warped Fermi cylinder to the SdH oscillations.
This has to be taken into account in the analysis by adding the beat factor $R_{\mathrm{b}}$ to the Lifshitz-Kosevich expression for the oscillation amplitude:
\begin{equation}\label{ABeats}
\frac{A}{B^{1/2} R_T} = A_0 R_{\mathrm{D}} R_{\mathrm{MB}} R_{\mathrm{b}}\,,
\end{equation}
where $A_0$ is a constant prefactor. In this equation we grouped the known quantities: the measured SdH amplitude, the magnetic field, and temperature damping factor on the left-hand side ($R_T$ contains, besides the fundamental constants, the known temperature and cyclotron mass, which has been determined independently).
Further, in our evaluations we assumed that $R_{\mathrm{MB}}=1$, i.e. $B_{\mathrm{MB}}=0$, see Eq.\,(\ref{RMB}). A comparison between Eqs.\,(\ref{RMB}) and (\ref{RD}) shows that substituting a true magnetic-breakdown field $B_{\mathrm{MB}} \sim 1$\,T (see Sec. II above) would result in a decrease of the estimated Dingle temperature by $\sim 0.03$\,K, which is small compared to the error bar of our estimations, see Fig.\,5 of the main text. Moreover, this correction is not expected to change significantly within our narrow pressure range $\sim 100$\,MPa. Thus, our study of the pressure dependence of the Dingle temperature should not be affected by the present assumption.

Coming to the beat factor $R_{\mathrm{b}}$,
when the interlayer bandwidth $4t_{\perp}$ strongly exceeds the Landau level spacing $\hbar \omega_c (= \frac{\hbar e B}{m_c})$, the oscillation amplitude is modulated as $R_{\mathrm{b}} = \cos \left( 2\pi \frac{F_{\mathrm{b}}}{B} - \frac{\pi}{4}\right)$ \cite{cham01a,kart04}.
The beat frequency, $F_{\mathrm{b}} = 2t_{\perp}m_c/\hbar e$, is easy to evaluate from the distance between the subsequent beat nodes.
However, in our experiment only one beat node was observed within the field range where the oscillations were resolved ($10\mathrm{\,T}-15$\,T), which makes the analysis more complicated. In particular, this indicates
that $t_{\perp} \sim \hbar \omega_c$ at these fields.
In this case the beats of the SdH signal acquire an additional, field-dependent phase $\phi(B)$ \cite{grig02b,grig03,mogi18}.
The existing experimental data on another layered organic metal \cite{grig02b} suggests that $\phi$ is proportional to the field:
\begin{equation}\label{Rb1}
R_{\mathrm{b}} = \cos \left( 2\pi \frac{F_{\mathrm{b}}}{B} - \frac{\pi}{4} + \alpha B \right).
\end{equation}
A detailed theory of the SdH effect in the interlayer conductivity of a metal with $t_{\perp} \sim \hbar \omega_c$ proposes a beat factor in the form \cite{grig03,mogi18}:
\begin{equation}\label{Rb-18}
R_{\mathrm{b}} = J_0\left( \frac{4\pi t_{\perp}}{\hbar \omega_c} \right) - \frac{\hbar \omega_c}{2\pi t_{\perp}} \left[ 1 + \frac{2\pi^2 k_{\mathrm{B}}T_{\mathrm{D}}}{\hbar \omega_c}J_1\left(  \frac{4\pi t_{\perp}}{\hbar \omega_c} \right) \right]\,,
\end{equation}
where $J_0(Z)$ and $J_1(Z)$ are 0-th and 1-st order Bessel functions, respectively.
This expression can be approximated to the form of Eq.\,(\ref{Rb1}) when $2\pi t_{\perp} > \hbar \omega_c > 2\pi^2 k_{\mathrm{B}}T_{\mathrm{D}}$, with $\alpha \approx \hbar \omega_c/2\pi t_{\perp}$. While the approximately linear $\phi(B)$ dependence looks consistent with the experimental data, the theory fails to reproduce the slope observed in the experiment \cite{grig02b}. Having in mind this uncertainty in understanding the phase shift of the beats, we fitted our data, using both the empirical Eq.\,(\ref{Rb1})
and the rigorous theoretical formula of Eq.\,(\ref{Rb-18}) for $R_{\mathrm{b}}$ Eq.\,(\ref{ABeats}).

Finally, at present we cannot disregard the possibility that the sample is composed of two slightly misaligned domains contributing to the SdH signal with slightly different frequencies, $F_{1,2} = F_{\beta} \pm \delta F$. In this case the beat factor is simply:
\begin{equation}\label{Rb3}
R_{\mathrm{b}} = \cos \left( 2\pi \frac{F_{\mathrm{b}}}{B}\right)
\end{equation}
with $F_{\mathrm{b}} = 2\delta F$.


Fig.\,\ref{TD-plots}(b) illustrates three different fits of the experimental data  with Eq.\,(\ref{ABeats}) using the beat factor in the form of Eq.\,(\ref{Rb1}), (\ref{Rb-18}), and (\ref{Rb3}), respectively.
The fitting parameters were: the amplitude prefactor $A_0$, $T_{\mathrm{D}}$ (entering the Dingle factor $R_{\mathrm{D}}$ and Eq.\,(\ref{Rb-18})), the beat frequency $F_{\mathrm{b}}$ (in Eq.\,(\ref{Rb-18}) it enters through the relation: $\frac{4\pi t_{\perp}}{\hbar \omega_c} = \frac{2\pi F_{\mathrm{b}}}{B}$), and, when using Eq.\,(\ref{Rb1}), the coefficient $\alpha$.
All three fits show a rather good agreement with the experiment although some deviations are observed at the edges of the field window when using Eqs.\,(\ref{Rb-18}), and (\ref{Rb3}). Very similar results have been obtained for all the other pressures in the range 27 to 99\,MPa. Fitting the 21\,MPa curve did not give a reliable result: because of the low oscillation amplitude at this pressure we were unable to locate the position of the beat node with a reasonable accuracy, which is an important component of the fitting procedure.
While using the empirical Eq.\,(\ref{Rb1}) yields a better fitting of the experimental data, it is not yet clear whether this is not simply because of using one variable parameter more than when using the other two equations for $R_{\mathrm{b}}$. Clarifying this is a matter of a separate study. At present we can conclude that all the performed fits yield similar values for the Dingle temperature, differing from each other within $\pm 0.1$\,K. This uncertainty gives the dominant contribution to the error bar of our estimations and is included in the error bars in Fig.\,5 of the main text.

The fitting procedure described above has yielded very similar $T_{\mathrm{D}}$ values for all pressures except the highest one. This can also be illustrated in a simple way by presenting the oscillation amplitude in the form of a ``Dingle plot'', see Fig.\,\ref{TD-plots}(c).
Without beating, the quantity $\ln(\frac{A_{\beta}}{B^{1/2}R_T})/\mu_c$ is a linear function of $1/B$ ($\mu_c \equiv m_c/m_0$); its slope $S$ solely depends on the Dingle temperature: $S = \frac{2\pi^2 m_0 k_{\mathrm{B}}}{\hbar e}T_{\mathrm{D}} \approx 14.7 T_{\mathrm{D}}$. In the presence of beats one can determine $S$ from the tangent to the Dingle plot provided the field window includes at least two antinodes. In our case the $1/B$ window, where the oscillations are observed, is just comparable but not larger than one half of the beat period, $\Delta_{\mathrm{b}}/2 = 1/(2 F_{\mathrm{b}}) \simeq 0.03$\,T$^{-1}$ [the latter is obtained from fitting the oscillation amplitude with Eq.\,(\ref{ABeats}) as described above]. At such conditions we cannot make a precise evaluation of $S$ in this way. Nevertheless, a comparison between the Dingle plots for different pressures can be used in order to check whether a significant variation of $S$ takes place. Fig.\,\ref{TD-plots}(c) clearly demonstrates that for all the pressures except the highest one (the top curve), the tangents to the Dingle plots have very similar slopes, in the range 13.6 to 16.3\,T , which would correspond to the Dingle temperatures all falling in a narrow interval $\approx0.9 - 1.1$\,K. Given the roughness of the present estimation, these values are in good agreement with the results of explicit fitting.

Finally, we address the apparent steplike drop of $T_{\mathrm{D}}$ at the highest pressure as presented in Fig.\,5 in the main text. To check whether this drop has an intrinsic physical meaning
we have carried out additional test runs in a broader pressure range including higher pressures. The measurements were done on the same $\kappa$-Cl sample as presented in the main text (sample \#1) at 0.4\,K. The results are shown by squares in Fig.\,\ref{TD-plots}(d) along with the data from Fig.\,5 of the main text (circles).
One can see that the point at $P=99$\,MPa clearly falls out of the remaining dataset. We, thus, disregard it as a spurious effect. This could be caused, for example, by an irreversible sample degradation after the measurements at this pressure, which were in fact the very first measurement cycle in our experiments.

\end{document}